\documentclass[letterpaper,12pt,notoc]{JHEP3}
\pdfoutput=1
\def\pd{\partial}
\def\Slash{{\!\!\!\!/}}

\usepackage{graphicx}
\usepackage{amsmath}
\usepackage{amssymb}

\preprint{ \hbox{}\hfill arXiv:1107.3413}

\title{Gravitational and Yang-Mills instantons in holographic RG flows}

\author{Edi Gava$^a$, Parinya Karndumri$^{a,\, b}$ and K. S. Narain$^c$\\
$^a$INFN, Sezione di Trieste, Italy\\
$^b$International School for Advanced Studies (SISSA), via Bonomea
265, 34136 Trieste, Italy \\
$^c$The Abdus Salam International Centre for Theoretical Physics,
Strada Costiera 11, 34100 Trieste, Italy \\
E-mail: \email{gava$@$ictp.it}, \email{karndumr$@$sissa.it},
\email{narain$@$ictp.it}}

\abstract{We study various holographic RG flow solutions involving
warped asymptotically locally Euclidean (ALE) spaces of $A_{N-1}$
type. A two-dimensional RG flow from a UV (2,0) CFT to a (4,0) CFT
in the IR is found in the context of (1,0) six dimensional
supergravity, interpolating between $AdS_3\times S^3/\mathbb{Z}_N$
and $AdS_3\times S^3$ geometries. We also find solutions involving
non trivial gauge fields in the form of $SU(2)$ Yang-Mills
instantons on ALE spaces. Both flows are of vev type, driven by a
vacuum expectation value of a marginal operator. RG flows in four
dimensional field theories are studied in the type IIB and type I$'$
context. In type IIB theory, the flow interpolates between
$AdS_5\times S^5/\mathbb{Z}_N$ and $AdS_5\times S^5$ geometries. The
field theory interpretation is that of an $N=2$ $SU(n)^N$ quiver
gauge theory flowing to $N=4$ $SU(n)$ gauge theory. In type I$'$
theory the solution describes an RG flow from $N=2$ quiver gauge
theory with a product gauge group to $N=2$ gauge theory in the IR,
with  gauge group $USp(n)$. The corresponding geometries are
$AdS_5\times S^5/(\mathbb{Z}_N\times \mathbb{Z}_2)$ and $AdS_5\times
S^5/\mathbb{Z}_2$, respectively. We also explore more general RG
flows, in which both the UV and IR CFTs are $N=2$ quiver gauge
theories and the corresponding geometries are $AdS_5\times
S^5/(\mathbb{Z}_N\times \mathbb{Z}_2)$ and $AdS_5\times
S^5/(\mathbb{Z}_M\times \mathbb{Z}_2)$. Finally, we discuss the
matching between the geometric and field theoretic pictures of the
flows.}

\begin{document}
\section{Introduction}
Holographic RG flow between D-dimensional (D=2,4) CFT's is one of
the most studied aspects of the AdS/CFT correspondence
\cite{maldacena}. The issue has been addressed both in the framework
of D+1 (possibly gauged) supergravity \cite{fgpw, an, gir, bs, gkn,
AP}, and in the full 10-dimensional supergravity, where the internal
space is typically a warped non-compact Calabi-Yau manifold
\cite{chethan, km}.
\\
\indent In this paper, we study RG flows in both two- and four-
dimensional contexts. In the previous paper \cite{rg_int}, we have
studied RG flows in minimal six dimensional supergravity with
Yang-Mills instantons turned on on $\mathbb{R}^4$. Here we still
work in the framework of (1,0) six dimensional supergravity, but now
we replace the transverse $\mathbb{R}^4$ with an ALE manifold of
$A_{N-1}$ type. We will adopt on it the well-known Gibbons-Hawking
multi-center metric \cite{Hawking ALE}.
\\
\indent Furthermore, we also study a flow solution involving
Yang-Mills instantons turned on on the ALE space, thereby
generalizing the solution discussed in \cite{rg_int}. Explicit
instanton solutions on an ALE space can be written down for the
$SU(2)$ gauge group \cite{Massimo}, \cite{SU2_instOnALE},
\cite{etesi} and we will then restrict ourselves to these solutions.
The resulting supergravity solutions describe RG flows in two
dimensional dual field theories and have asymptotic geometries
$AdS_3\times S^3/\mathbb{Z}_N$ in the UV and $AdS_3\times S^3$ in
the IR. The former arises from the limit where one goes to the
boundary of the ALE, the latter when one zooms near one of the
smooth ALE centers. Notice that in this case the solution describes
the flow from a (2,0) UV CFT to a (4,0) IR CFT, contrary to the case
of \cite{rg_int}, where both fixed points were (4,0) CFT's. Indeed,
in the UV we have  $\mathbb{Z}_N$ projection, due to asymptotic
topology of the ALE space.
\\
\indent We will then move to study flow solutions in 10D type IIB
and type I$'$ theories (by the latter we mean IIB on
$T^2/(-1)^{F_L}\Omega I_2$, the double T-dual of type I on $T^2$
\cite{Sen_K3F_theory}) on an ALE background. These solutions
describe RG flows of four dimensional UV CFT's with $N=2$
supersymmetry. In the type IIB case our solution is a variation on
the theme discussed in \cite{chethan, km} for the ALE space of the
form $\mathbb{C}^3/\mathbb{Z}_3$ and for the conifold, respectively,
which describe flows from $N=1$ to $N=4$ CFT's. The RG flow in type
IIB theory on $\mathbb{C}^3/\mathbb{Z}_3$ has also been studied in
more details in \cite{nessi}, recently. Our flows interpolate
between $N=2$ quiver gauge theories with product gauge group in the
UV and the $N=4$ $SU(n)$ supersymmetric Yang-Mills theory in the IR.
The corresponding asymptotic geometries are $AdS_5\times
S^5/\mathbb{Z}_N$ and $AdS_5\times S^5$.
\\
\indent The discussion becomes more interesting in type I$'$ theory:
in this case we find that the critical points are described by the
geometries $AdS_5\times S^5/(\mathbb{Z}_N\times \mathbb{Z}_2)$ in
the UV and $AdS_5\times S^5/ \mathbb{Z}_2$ in the IR. The
$\mathbb{Z}_2$ is identified with $(-1)^{F_L}\Omega I_2$. The UV
gauge groups are more complicated than those of type IIB case, and
are among the (unoriented) quiver gauge groups discussed in
\cite{douglas_quiver}. The quiver diagrams have different structures
depending on whether $N$ is even or odd, and for $N$ even there are
in addition two possible projections, resulting in two different
quiver structures. This is what will make the discussion of RG flows
richer and more interesting. We will in fact verify the agreement
between the geometric picture emerging from the supergravity
solutions and the corresponding field theory description, where the
flows are related to the Higgsing of the gauge group, i.e. they are
driven by vacuum expectation values of scalar fields belonging to
the hypermultiplets of the $N=2$ theories.
\\
\indent We also consider more general RG flows, in which not all the
UV gauge group is broken to a single diagonal IR subgroup. In other
words, the IR theory can be another, smaller, quiver gauge theory.
The associated flows are the flows between two $N=2$ quiver gauge
theories, and the corresponding geometries are given by $AdS_5\times
S^5/(\mathbb{Z}_N\times \mathbb{Z}_2)$ and $AdS_5\times
S^5/(\mathbb{Z}_M\times \mathbb{Z}_2)$ with $M<N$. We will find that
field theory considerations do not allow all possible flows with
arbitrary values of $M$ and $N$ and some symmetry breaking patterns
are forbidden. Actually, we will see that these features are
reproduced by the geometry, having to do with the fact that the
$\Omega$ projection  does not allow arbitrary ALE geometry since it
projects out or identify the geometric moduli, as was already
observed in a different context in \cite{pol_K3Orientifold}. In
fact, we will obtain a very satisfactory agreement between with the
field theory and the supergravity pictures.
\\
\indent The paper is organized as follows. In section \ref{2Dflow},
we present flow solutions in (1,0) six dimensional supergravity
coupled to an anti-symmetric tensor multiplet on the ALE background.
We then add $SU(2)$ instantons by coupling the supergravity theory
to $SU(2)$ Yang-Mills multiplets and turning on the instantons on
the ALE space. In section \ref{4Dflow}, we find supersymmetric
solutions to type IIB and type I$'$ theories. Unfortunately, in this
case, we are not able to obtain the explicit form of the solutions.
However their existence, with the required boundary conditions are
guaranteed on general mathematical grounds. The central charges
along with field theory descriptions of the flows are also given. In
section \ref{geometry}, we consider more general RG flows in which
both the UV and IR CFTs are $N=2$ quiver gauge theories and give a
geometric interpretation for the symmetry breaking patterns.
Finally, we make some conclusions and comments in section
\ref{conclusion}.
\section{RG flows in six dimensional supergravity} \label{2Dflow}
In this section, we will find flows solution in (1,0) six
dimensional supergravity. We begin with a review of (1,0)
supergravity and focus mainly on relevant formulae we will use
throughout this section. We proceed by studying an RG flow solution
on the ALE background and compute the ratio of the central charges
of the UV and IR fixed points. We then include $SU(2)$ instantons on
the ALE background. This is also a generalization of the solution
studied in \cite{rg_int} in which the flow involves only Yang-Mills
instantons. We will see that the result is a combined effect of
gravitational instantons studied here and $SU(2)$ instantons studied
in \cite{rg_int}. Finally, we discuss the left and right central
charges with a subleading correction including curvature squared
terms on the gravity side.
\subsection{An RG flow with graviational instantons}
We now study a supersymmetric RG flow solution in (1,0) six
dimensional supergravity constructed in \cite{nishino}. We are
interested in the six dimensional supergravity theory coupled to one
tensor multiplet and $SU(2)$ Yang-Mills multiplets. We refer the
reader to \cite{nishino} for the detailed construction of this
theory. The equations of motion for bosonic fields are given by
\cite{nishino}
\begin{eqnarray}
\hat{R}_{MN}-\frac{1}{2}\hat{g}_{MN}\hat{R}-\frac{1}{3}e^{2\hat{\theta}}\left(3{\hat{G}}_{3MPQ}{\hat{G}}_{3N}^{\phantom{asa}PQ}
-\frac{1}{2}\hat{g}_{MN}{\hat{G}}_{3PQR}{\hat{G}_3}^{PQR}\right)\nonumber
\\-\partial_M\hat{\theta}
\partial^M\hat{\theta}+\frac{1}{2}\hat{g}_{MN}\partial_P\hat{\theta}\partial^P\hat{\theta}-e^{\hat{\theta}}
\left(2\hat{F}^{IP}_M\hat{F}^I_{NP}-\frac{1}{2}\hat{g}_{MN}\hat{F}^I_{PQ}\hat{F}^{IPQ}\right)&=&0,\\
\hat{D}(e^{2\hat{\theta}}\hat{*}\hat{G}_3)+\tilde{v} \hat{F}^I\wedge \hat{F}^I&=&0,\label{G3eq1}\\
\hat{D}[(ve^{\hat{\theta}}+\tilde{v}e^{-\hat{\theta}})
\hat{*}\hat{F}^I]-2ve^{2\hat{\theta}}\hat{*}\hat{G}_3\wedge
\hat{F}^I+2\tilde{v}\hat{*}\hat{G}_3\wedge
\hat{F}^I&=&0,\label{FIeq1} \\
\hat{d}\hat{*}\hat{d}\hat{\theta}+(ve^{\hat{\theta}}+\tilde{v}e^{-\hat{\theta}})
\hat{*}\hat{F}^I\wedge
\hat{F}^I+2e^{2\hat{\theta}}\hat{*}\hat{G}_3\wedge\hat{G}_3&=&0\label{THeq1}
\end{eqnarray}
with the Bianchi identity for $\hat{G}_3$ given by
\begin{equation}
\hat{D}\hat{G}_3=v\hat{F}^I\wedge \hat{F}^I\, . \label{Bianchi}
\end{equation}
$\hat{\theta}$ is the scalar field in the tensor multiplet. Indices
$M, N, \ldots=0,\ldots, 5$ label six dimensional coordinates, and
$I, J,\ldots$ are adjoint indices of the corresponding Yang-Mills
gauge group, $SU(2)$ in the present case. For both $v$ and
$\tilde{v}$ non-zero, there is no invariant Lagrangian, but the
presence of the Lagrangian is not relevant for our discussion. As in
\cite{rg_int}, we also assume that both $v$ and $\tilde{v}$ are
positive, and the hatted fields are six-dimensional ones. We also
need supersymmetry transformations of fermionic fields which, in
this case, are the gravitino $\psi_M$, gauginos $\lambda^I$ and the
fermion in the tensor multiplet $\chi$. With fermions being zero,
these transformations are given by \cite{nishino}
\begin{eqnarray}
\delta \psi_M &=& \hat{D}_M \epsilon +\frac{1}{24}e^{\hat{\theta}}
\Gamma^{NPQ}\Gamma_M \hat{G}_{3NPQ}\epsilon, \label{deltapsi}\\
\delta \lambda^I&=&\frac{1}{4}\Gamma^{MN}\hat{F}^I_{MN}\epsilon,
\label{deltalambda}
\\
\delta \chi&=&\frac{1}{2}\Gamma^M\partial_M\hat{\theta}\epsilon
-\frac{1}{12}e^{\hat{\theta}}\Gamma^{MNP}\hat{G}_{3MNP}\epsilon\, .
\label{deltachi}
\end{eqnarray}
The metric ansatz is
\begin{eqnarray}
ds^2_6=e^{2f}(-dx_0^2+dx^2_1)+e^{2g}ds^2_4\,
.\label{6DMetric_ansatz}
\end{eqnarray}
The four dimensional metric $ds^2_4$ will be chosen to be the
gravitational multi-instantons of \cite{Hawking ALE}. This is an
asymptotically locally Euclidean space (ALE) with the metric
\begin{equation}
ds^2_4=V^{-1}(d\tau+\vec{\omega}.d\vec{x})^2+Vd\vec{x}.d\vec{x}.\label{ALEmetric}
\end{equation}
The function $V$ is given by
\begin{equation}
V=\sum_{i=1}^N\frac{1}{|\vec{x}-\vec{x}_i|}\, .
\end{equation}
The function $\vec{\omega}$ is related to $V$ via
\begin{equation}
\vec{\nabla}\times \vec{\omega}=\vec{\nabla} V,
\end{equation}
and the $\tau$ has period $4\pi$. We also choose the gauge
\begin{equation}
\vec{\omega}.d\vec{x}=\sum_{i=1}^N\cos\theta_id\phi_i
\end{equation}
as in \cite{Page}. The point $\vec{x}_i$ is the origin of the
spherical coordinates $(r_i,\theta_i,\phi_i)$ with
$r_i=|\vec{x}-\vec{x}_i|$. The procedure is now closely parallel to
that of \cite{rg_int}, so we only repeat the main results here and
refer the reader to \cite{rg_int} for the full derivation. Although
$\hat{A}^I=0$ in the present case, it is more convenient to work
with non-zero $\hat{A^I}$ since equations with non-zero $\hat{A}^I$
will be used later in the next subsection. We will keep
$ds^2_4=g_{\alpha\beta}dz^\alpha dz^\beta$, $\alpha=2,3,4,5$, in
deriving all the necessary equations. The ansatz for $\hat{G}_3$ and
$\hat{A}^I$, $I=1,2,3$ are
\begin{equation}
\hat{A}^I=A^I , \qquad \hat{F}^I=F^I, \qquad \hat{G}_3=G+dx_0\wedge
dx_1\wedge d\Lambda \, .
\end{equation}
As in \cite{rg_int}, unhatted fields are four dimensional ones
living on the four dimensional space with the metric $ds^2_4$ and
depending only on $z^\alpha$. Equation \eqref{G3eq1} gives
\begin{eqnarray}
D(e^{2\theta-2f+2g}*d\Lambda)&=&vF^I\wedge F^I,\label{stardLambdaeq}\\
*G&=&e^{-2\theta-2f+2g}d\tilde{\Lambda}\label{starG}
\end{eqnarray}
where $\tilde{\Lambda}$ is a $z^\alpha$ dependent function. We now
take $F^I$ to be self dual with respect to four dimensional $*$. In
order to solve the Killing spinor equations, we impose the chirality
condition $\Gamma_{01}\epsilon=\epsilon$ which implies
$\Gamma_{2345}\epsilon=\epsilon$ by the six dimensional chirality
$\Gamma_7\epsilon=\epsilon$. So, equation $\delta\lambda^I=0$ is
trivially satisfied because $\Gamma_{\alpha\beta}$ is anti-selfdual.
Furthermore, the condition $\Gamma_{01}\epsilon=\epsilon$ also
breaks half of the (1,0) supersymmetry, so our flow solution
preserves half of the eight supercharges. It is easy to show that
$\delta\chi=0$ and $\delta\psi_\mu=0$ give, respectively,
\begin{eqnarray}
\pd\Slash\, \theta+e^{\theta-2f}\pd\Slash \Lambda -e^{-\theta-2f}\pd
\Slash \tilde{\Lambda} &=&0,
\label{deltaChi} \\
\Gamma_\mu \pd \Slash \, f-\frac{1}{2}e^{\theta-2f}\Gamma_\mu \pd
\Slash \, \Lambda -\frac{1}{2}e^{-\theta-2f}\Gamma_\mu\pd \Slash
\tilde{\Lambda}&=&0\, . \label{deltaPsiMu}
\end{eqnarray}
Taking combinations $\eqref{deltaChi}\pm \eqref{deltaPsiMu}$, we
find
\begin{equation}
\Lambda=\frac{1}{2}e^{-\theta+2f}+C_1,\qquad
\tilde{\Lambda}=\frac{1}{2}e^{\theta+2f}+C_2\label{Lambda}
\end{equation}
with constants of integration $C_1$ and $C_2$. Equation
$\delta\psi_\alpha=0$ reads
\begin{equation}
D_\alpha\tilde{\epsilon}-\frac{1}{2}\Gamma_{\beta\alpha}\pd^\beta(f+g)\tilde{\epsilon}=0
\, .\label{internalGravitino}
\end{equation}
We have used $\epsilon=e^{\frac{f}{2}}\tilde{\epsilon}$. Equation
\eqref{internalGravitino} can be satisfied provided that $g=-f$ and
\begin{equation}
D_\alpha \tilde{\epsilon}=0\, .
\end{equation}
The latter condition requires that $\tilde{\epsilon}$ is a Killing
spinor on the ALE space. The ALE space has $SU(2)$ holonomy and
admits two Killing spinors out of the four spinors. Therefore, the
flow solution entirely preserves $\frac{1}{4}$ of the eight
supercharges, or $N=2$ in two dimensional langauge, along the flow.
\\
\indent In this subsection, we study only the effect of
gravitational instantons, so we choose $A^I=0$ from now on. Using
\eqref{Lambda}, we can write \eqref{Bianchi} and
\eqref{stardLambdaeq} as
\begin{equation}
\square\, e^{-\theta-2f}=0\qquad \textrm{and} \qquad \square\,
e^{\theta-2f}=0\, .
\end{equation}
The $\square$ in these equations is the covariant scalar Laplacian
on the ALE space
\begin{equation}
\square=\frac{1}{V}[V^2\pd^2_\tau+(\vec{\nabla}-\vec{\omega}\pd_\tau).(\vec{\nabla}-\vec{\omega}\pd_\tau)].
\end{equation}
Our flow is described by a simple ansatz as follows. We first choose
$\theta=0$. It is straightforward to check that all equations of
motion as well as BPS equations are satisfied. We then have only a
single equation to be solved
\begin{equation}
\square\,e^{-2f}=0\, .\label{boxf2}
\end{equation}
We now choose $f$ to be $\tau$ independent of the form
\begin{equation}
e^{-2f}=\frac{c}{|\vec{x}-\vec{x_1}|}\label{ansatz1}
\end{equation}
where $c$ is a constant. This is clearly a solution of \eqref{boxf2}
since for $\tau$ independent functions, the $\square$ reduce to the
standard three dimensional Laplacian $\vec{\nabla}.\vec{\nabla}$. We
will now show that this solution describes an RG flow between two
fixed points given by $|\vec{x}|\rightarrow \infty$ and
$\vec{x}\rightarrow \vec{x}_1$. We emphasize that the point
$\vec{x}_1$ is purely conventional since any point $x_i$ with
$i=1\ldots N$ will work in the same way. Notice that for general
$\tau$ dependent solution, the solution to the harmonic function
will be given by the Green function on ALE spaces. The explicit form
of this Green function will be given in the next subsection.
Furthermore, with $\tau$ dependent solution, the IR fixed point of
the flow can also be given by $\vec{x}\rightarrow \vec{y}$ where
$\vec{y}$ is a regular point on the ALE space rather than one of the
ALE canter $\vec{x}_i$. The crucial point in our discussion is the
behavior of the Green function near the fixed points such that the
geometry contains $AdS_3$. However, for the present case, we
restrict ourselves to the ansatz \eqref{ansatz1}.
\\
\indent When $|\vec{x}|\rightarrow \infty$, we have
\begin{eqnarray}
e^{-2f}&=&\frac{c}{|\vec{x}-\vec{x}_1|}\rightarrow
\frac{c}{\zeta},\qquad \zeta \equiv|\vec{x}|,\nonumber \\
V&=&\sum_{i=1}^N\frac{1}{|\vec{x}-\vec{x}_i|}\rightarrow
\frac{N}{\zeta}\, .
\end{eqnarray}
In this limit, the ALE metric becomes
\begin{equation}
ds^2_4=\frac{\zeta}{N}(d\tau+N\cos\theta
d\phi)^2+\frac{N}{\zeta}(d\zeta^2+\zeta^2 d\Omega_2^2)
\end{equation}
where we have written the flat three dimensional metric
$d\vec{x}.d\vec{x}$ in spherical coordinates with the $S^2$ metric
$d\Omega^2_2$. The factor $N\cos\theta d\phi$ arises from
$\sum_{i=1}^N\cos\theta_id\phi_i$ since in the limit
$|\vec{x}|\rightarrow \infty$ all $(\theta_i,\phi_i)$ are the same
to leading order. By changing the coordinate $\zeta$ to $r$ defined
by $\zeta=\frac{r^2}{4N}$, we obtain
\begin{equation}
ds^2_4=dr^2+\frac{r^2}{4}\left[\left(\frac{d\tau}{N}+\cos\theta
d\phi\right)^2+d\Omega^2_2\right].
\end{equation}
The full six-dimensional metric is then given by
\begin{equation}
ds^2_6=\frac{r^2}{4Nc}dx^2_{1,1}+\frac{4Nc}{r^2}dr^2+4Nc\left[\left(\frac{d\tau}{N}+\cos\theta
d\phi\right)^2+d\Omega^2_2\right].
\end{equation}
The expression in the bracket is the metric on $S^3/\mathbb{Z}_N$.
So, the six dimensional geometry is $AdS_3\times S^3/\mathbb{Z}_N$
with the radii of $AdS_3$ and $S^3/\mathbb{Z}_N$ being $L_\infty
=2\sqrt{Nc}$.
\\
\indent When $\vec{x}\rightarrow \vec{x}_1$, we find
\begin{eqnarray}
e^{-2f}&=&
\frac{c}{\xi},\qquad \xi \equiv|\vec{x}-\vec{x}_1|,\nonumber \\
V&=&\sum_{i=1}^N\frac{1}{|\vec{x}-\vec{x}_i|}= \frac{1}{\xi}\, .
\end{eqnarray}
The ALE metric becomes
\begin{equation}
ds^2_4=\xi (d\tau+\cos\theta_1
d\phi_1)^2+\frac{1}{\xi}(d\xi^2+\xi^2d\Omega_2^2).
\end{equation}
In this limit, $\sum_{i=1}^N\cos\theta_id\phi_i\sim
\cos\theta_1d\phi_1$ to leading order. Writing $\xi=\frac{r^2}{4}$,
we obtain
\begin{equation}
ds^2_4=dr^2+\frac{r^2}{4}[(d\tau-\cos \theta_1
d\phi_1)^2+d\Omega_2^2]
\end{equation}
which is the metric on $\mathbb{R}^4$. The six-dimensional metric
now takes the form
\begin{equation}
ds^2_6=\frac{r^2}{4c}dx_{1,1}^2+\frac{4c}{r^2}dr^2+4c d\Omega^2_3
\end{equation}
where $d\Omega^2_3$ is the metric on $S^3$. This geometry is
$AdS_3\times S^3$ with $AdS_3$ and $S^3$ having the same radius
$2\sqrt{c}$. The central charge of the dual CFT is given by
\begin{equation}
c=\frac{3L}{2G_N^{(3)}}\, .
\end{equation}
We find the ratio of the central charges
\begin{eqnarray}
\frac{c_1}{c_\infty}&=&\frac{L_1G^{(3)}_{N\infty}}{L_\infty
G^{(3)}_{N1}}=\frac{L_1\textrm{Vol}(S^3)}{L_\infty
\textrm{Vol}(S^3/\mathbb{Z}_N)} \nonumber \\
&=&N\left(\frac{L_1}{L_\infty} \right)^4=\frac{1}{N}
\end{eqnarray}
where we have used $G^{(3)}_N=\frac{G_N^{(6)}}{\textrm{Vol}(M)}$ for
six-dimensional theory compactified on a compact space $M$. The flow
interpolates between $AdS_3\times S^3/\mathbb{Z}_N$ in the UV to
$AdS_3\times S^3$ in the IR. The UV CFT has (2,0) supersymmetry
because of the $\mathbb{Z}_N$ projection, so our flow describes an
RG flow from the (2,0) CFT in the UV to the (4,0) CFT in the IR.
\\
\indent We now consider the central charge on the gravity side
including the curvature squared terms. The bulk gravity is three
dimensional, and the Riemann tensor can be written in terms of the
Ricci tensor and Ricci scalar. To study the effect of higher
derivative terms, we add the
$R_{\mu\nu\rho\sigma}R^{\mu\nu\rho\sigma}$ term to the (1,0) six
dimensional action. The supersymmetrization of this term has been
studied in \cite{bergshoeff}. We temporarily drop the hat to
simplify the expressions. The Lagrangian with the auxiliary fields
integrated out is given by \cite{pope_massive_3D}
\begin{eqnarray}
\mathcal{L}&=&\sqrt{-g}e^{-2\theta}\left[R+4\pd_\mu \theta\pd^\mu
\theta-\frac{1}{12}G_3^{\mu\nu\rho}G_{3\mu\nu\rho}\right]+\frac{1}{4}\alpha
\sqrt{-g}\tilde{R}^{\mu\nu\rho\sigma}\tilde{R}_{\mu\nu\rho\sigma}\nonumber
\\
& &+\frac{1}{16}\beta
\epsilon^{\mu\nu\rho\sigma\tau\lambda}\tilde{R}^{\alpha\beta}_{\phantom{sa}\mu\nu}
\tilde{R}_{\alpha\beta\rho\sigma}b_{\tau\lambda}\label{6DR2action}
\end{eqnarray}
where $\tilde{R}_{\mu\nu\rho\sigma}$ is computed with the modified
connection
$\tilde{\Gamma}^\rho_{\phantom{s}\mu\nu}=\Gamma^\rho_{\phantom{s}\mu\nu}-\frac{1}{2}G^\rho_{3\mu\nu}$.
The $b_{\lambda\tau}$ is the two-form field whose field strength is
$G_3$. Reducing \eqref{6DR2action} on $S^3$ with
$G_3=2S\epsilon_3+2m\omega_3$ where $\epsilon_3$ and $\omega_3$ are
volume forms on $ds^2_3$ and $S^3$, respectively gives
\cite{pope_massive_3D}
\begin{eqnarray}
e^{-1}\mathcal{L}&=&e^{-2\theta}(R+4\pd_\mu \theta\pd^\mu
\theta+4m^2+2S^2)+4mS\nonumber \\
& &-2\beta
m\left[RS+2S^3-\frac{1}{4}\epsilon^{\mu\nu\rho}\left(R^{ab}_{\phantom{sa}\mu\nu}\omega_{\rho
ab}+\frac{2}{3}\omega_{\mu\phantom{a}b}^{\phantom{s}a}\omega_{\nu\phantom{a}c}^{\phantom{s}b}
\omega_{\rho\phantom{a}a}^{\phantom{s}c}\right)\right]\nonumber \\
& & +\frac{1}{4}\alpha (4R^{\mu\nu}R_{\mu\nu}-R^2-8\pd_\mu S\pd^\mu
S+12S^4+4RS^2).
\end{eqnarray}
As shown in \cite{pope_massive_3D}, $S=-m$ on the $AdS_3$
background, and $m$ is related to the AdS radius via
$m=\frac{1}{L}$. The left and right moving central charges can be
computed as in \cite{kraus, kraus_higherD}. The result is
\cite{pope_massive_3D}
\begin{equation}
c_L=\frac{3L}{2G^{(3)}_N}\left(1+\frac{4\beta}{L^2}\right),\qquad
c_R=\frac{3L}{2G^{(3)}_N}\, .
\end{equation}
We find that
\begin{eqnarray}
  \textrm{UV}:\qquad c_L&=&\frac{48\pi^2c^2N}{G^{(6)}_N}\left(1+\frac{\beta}{cN}\right),\qquad c_R=\frac{48\pi^2
  c^2N}{G_N^{(6)}},\\
  \textrm{IR}:\qquad c_L&=&\frac{48\pi^2c^2}{G^{(6)}_N}\left(1+\frac{\beta}{c}\right),\qquad c_R=\frac{48\pi^2
  c^2}{G_N^{(6)}}\, .
\end{eqnarray}
\indent We end this subsection by finding the dimension of the dual
operator driving the flow. This is achieved by expanding the metric
around the UV fixed point, $|\vec{x}|\rightarrow \infty$ in our
solution. $e^{-2f}$ and $V$ can be expanded as
\begin{eqnarray}
e^{-2f}&=&\frac{c}{|\vec{x}-\vec{x}_1|}\sim
\frac{1}{\zeta}\left(1+\frac{a_1\cos\varphi_1}{\zeta}-\frac{a_1^2}{2\zeta^2}\right)+\ldots,\nonumber
\\
V&=&\sum_{i=1}^N\frac{1}{|\vec{x}-\vec{x}_i|}\sim
\frac{N}{\zeta}+\sum_{i=1}^N\left(\frac{a_i\cos
\varphi_i}{\zeta^2}-\frac{a_i^2}{2\zeta^3}\right)+\ldots\label{expandsion}
\end{eqnarray}
where $\varphi_i$ are angles between $\vec{x}$ and $\vec{x}_i$. We
have also defined $\zeta\equiv |\vec{x}|$ and $a_i\equiv
|\vec{x}_i|$. By substituting \eqref{expandsion} in
\eqref{6DMetric_ansatz}, it is then straightforward to obtain the
behavior of the metric fluctuation which is of order
$\mathcal{O}(r^{-2})$. This gives $\Delta=2$ indicating that the
flow is driven by a vacuum expectation value of a marginal operator.
\subsection{An RG flow with gravitational and $SU(2)$ Yang-Mills
instantons} We now add Yang-Mills instantons to the solution given
in the previous subsection. This involves constructing instantons on
ALE spaces. Some explicit instantons solutions on an ALE space are
given in \cite{Massimo}. We are interested in $SU(2)$ instantons
whose explicit solutions can be written down. The solution can be
expressed in the form \cite{Massimo}
\begin{equation}
A^I_\alpha dx^\alpha=-\eta^I_{ab}e^aE^b\ln H\, .
\end{equation}
The vielbein $e^a_\alpha$ and its inverse $E^\alpha_a$ for the
metric \eqref{ALEmetric} are given by
\begin{eqnarray}
e^0&=&V^{-\frac{1}{2}}(d\tau+\vec{\omega}.d\vec{x}),\qquad
e^l=V^{\frac{1}{2}}dx^l, \\
E_0&=&V^{\frac{1}{2}}\frac{\pd}{\pd \tau}, \qquad
E_l=V^{-\frac{1}{2}}\left(\frac{\pd}{\pd x^l}-\omega_l\frac{\pd}{\pd
\tau}\right).
\end{eqnarray}
The $\eta^I_{ab}$'s are the usual 't Hooft tensors and $l=1,2,3$.
This form resembles the $SU(2)$ instantons on the flat space
$\mathbb{R}^4$. Self duality of $F^I$ requires that $H$ satisfies
the harmonic equation on the ALE space
\begin{equation}
\nabla_a\nabla^aH=0\, .
\end{equation}
The solution is given by
\begin{equation}
H=H_0+\sum_{j=1}^n\lambda_jG(x,y_j)\label{Hsol}
\end{equation}
where $H_0$ and $\lambda_j$ are constants, and $G(x,y_j)$ is the
Green's function on the ALE space given in \cite{Page} with
$x=(\tau,\vec{x})$. Its explicit form is
\begin{equation}
G(x,x')=\frac{\sinh U}{16\pi^2|\vec{x}-\vec{x}'|(\cosh U-\cos
T)}\label{PageG}
\end{equation}
where
\begin{eqnarray}
U(x,x')&=&\frac{1}{2}\sum_{i=1}^N\ln\left(\frac{r_i+r'_i+|\vec{x}-\vec{x}'|}{r_i+r'_i-|\vec{x}-\vec{x}'|}\right),
\qquad r_i=|\vec{x}-\vec{x}_i|,\nonumber \\
T(x,x')&=&\frac{1}{2}(\tau-\tau') +\sum_{i=1}^N\tan^{-1}
\left[\tan\left[\frac{\phi_i-\phi'_i}{2}\right] \frac{\cos
\frac{\theta_i+\theta'_i}{2}}{\cos\frac{\theta_i-\theta'_i}{2}}
\right].\label{UT_def}
\end{eqnarray}
This solution is obviously $\tau$ dependent and can be thought of as
a generalization of the $\tau$ independent solution of
\cite{SU2_instOnALE}. The latter is subject to the constraint $n\leq
N$ since the finite action requires that the instantons must be put
at the ALE centers. We emphasize here that the $\vec{y}_j$ inside
the $y_j$ in \eqref{Hsol} needs not necessarily coincide with the
ALE center $\vec{x}_i$. Therefore, $\vec{y}_j$ could be any point,
ALE center or regular point, on the ALE space. However, in our flow
solution given below, we will choose one of the $\vec{y}_j$'s to
coincide with one of the ALE centers $\vec{x}_i$'s which is, by our
convention, chosen to be $\vec{x}_1$.
\\
\indent As in the flat space case, we can write
\begin{equation}
F^I_{ab}F^{Iab}=-4\square\square \ln H
\end{equation}
which can be shown by using the properties of $\eta^I_{ab}$ given in
\cite{lecture_instanton} and the fact that $H$ is a harmonic
function on the ALE space as well as the Ricci flatness of the ALE
space. Using this relation, we obtain
\begin{equation}
*(F^I\wedge F^I)=*(*F^I\wedge
F^I)=\frac{1}{2}F^I_{ij}F^{Iij}=-2\square\square \ln H\, .
\end{equation}
Equations \eqref{Bianchi} and \eqref{stardLambdaeq} become
\begin{eqnarray}
\square \, e^{-\theta-2f}&=&4v\square\square \ln H,\\
\square \, e^{\theta-2f}&=&4\tilde{v}\square\square \ln H\, .
\end{eqnarray}
The solutions to these equations are of the form
\begin{eqnarray}
e^{-\theta-2f}&=&f_1+4v\square \ln H, \nonumber \\
e^{\theta-2f}&=&f_2+4\tilde{v}\square \ln H\label{sol1}
\end{eqnarray}
where $f_1$ and $f_2$ are solutions to the homogeneous equations.
The Green function $G(x,x')$ in \eqref{PageG} is singular when
$x\sim x'$. The behavior of $G(x,x')$ in this limit is \cite{Page}
\begin{equation}
G(x,x')=\frac{1}{4\pi^2|x-x'|^2}
\end{equation}
where
\begin{equation}
|x-x'|^2=V|\vec{x}-\vec{x}'|^2+V^{-1}[\tau-\tau'+\vec{\omega}.(\vec{x}-\vec{x}')]^2.
\end{equation}
We remove this singularity, in our case $x'\sim y_j$, from our
solution by adding $G(x,y_j)$, with appropriate coefficients, to
\eqref{sol1}. We also choose $f_1$ and $f_2$ to be
$\frac{c}{|\vec{x}-\vec{x}_1|}$ and $\frac{d}{|\vec{x}-\vec{x}_1|}$,
respectively. This choice is analogous to the solution in the
previous subsection with $c$ and $d$ being constants. Collecting all
these, we find
\begin{eqnarray}
e^{-\theta-2f}&=&\frac{c}{|\vec{x}-\vec{x}_1|}\nonumber
\\ & &+4v\left[\square \ln \left(H_0+\sum^n_{j=1}\lambda_j
G(x,y_j)\right)+16\pi^2\sum_{j=1}^n
G(x,y_j)\right], \label{sol2} \\
e^{\theta-2f}&=&\frac{d}{|\vec{x}-\vec{x}_1|}\nonumber
\\ & &+4\tilde{v}\left[\square
\ln \left(H_0+\sum^n_{j=1}\lambda_j
G(x,y_j)\right)+16\pi^2\sum_{j=1}^n G(x,y_j)\right].\label{sol3}
\end{eqnarray}
The metric warp factor $e^{-2f}$ can be obtained by multiplying
\eqref{sol2} and \eqref{sol3}. We now study the behavior of this
function in the limits $\vec{x}\rightarrow \vec{x}_1$ and
$|\vec{x}|\rightarrow \infty$.
\\
\indent As $\vec{x}\rightarrow \vec{x}_1$, the terms involving
$G(x,x_1)$ in the square bracket in \eqref{sol2} and \eqref{sol3} do
not contribute since the poles of the two terms cancel each other.
The other terms involving $G(x,y_j)$, $\vec{y}_j\neq \vec{x}_1$, are
subleading compared to $f_1$ and $f_2$. We find
\begin{equation}
e^{-\theta-2f}=\frac{d}{|\vec{x}-\vec{x}_1|},\qquad
e^{\theta-2f}=\frac{c}{|\vec{x}-\vec{x}_1|}
\end{equation}
or
\begin{equation}
e^{-2f}=\frac{\sqrt{cd}}{|\vec{x}-\vec{x_1}|}\, .
\end{equation}
By using the coordinate changing as in the previous subsection
$|\vec{x}-\vec{x}_1|=\frac{r^2}{4}$, it can be shown that the metric
is of the form of $AdS_3\times S^3$
\begin{equation}
ds^2_6=\frac{r^2}{4\sqrt{cd}}dx_{1,1}^2+\frac{4\sqrt{cd}}{r^2}dr^2+4\sqrt{cd}
d\Omega^2_3\, .
\end{equation}
\indent As $|\vec{x}|\rightarrow \infty$, the Green function
\eqref{PageG} becomes
\begin{equation}
G(x,x')=\frac{1}{16\pi^2|\vec{x}-\vec{x}'|}
\end{equation}
because $U$ defined in \eqref{UT_def} becomes infinite. We find
\begin{eqnarray}
e^{-\theta-2f}&=&\frac{c}{|\vec{x}-\vec{x}_1|}+4v\sum^n_{i=1}\frac{1}{|\vec{x}-\vec{y}_i|}\sim
\frac{c+4vn}{|\vec{x}|}, \nonumber \\
e^{\theta-2f}&=&\frac{d}{|\vec{x}-\vec{x}_1|}+4\tilde{v}\sum^n_{i=1}\frac{1}{|\vec{x}-\vec{y}_i|}\sim
\frac{d+4\tilde{v}n}{|\vec{x}|}\, .
\end{eqnarray}
The warp factor is now given by
\begin{equation}
e^{-2f}=\frac{\sqrt{(c+4nv)(d+4n\tilde{v})}}{|\vec{x}|}\, .
\end{equation}
The six-dimensional metric becomes $AdS_3\times S^3/\mathbb{Z}_N$,
with $|\vec{x}|=\frac{r^2}{4N}$,
\begin{equation}
ds^2_6=\frac{r^2}{\ell^2}dx^2_{1,1}+\frac{\ell^2}{r^2}dr^2+\ell^2\left[\left(\frac{d\tau}{N}+\cos\theta
d\phi\right)^2+d\Omega^2_2\right]
\end{equation}
where the $AdS_3$ radius is given by
\begin{equation}
\ell=2\sqrt{N}[(c+4nv)(d+4n\tilde{v})]^{\frac{1}{4}}\, .
\end{equation}
The ratio of the central charges can be found in the same way as
that in the previous subsection and is given by
\begin{equation}
\frac{c_1}{c_\infty}=N\left(\frac{L_1}{L_\infty}\right)^{4}=\frac{cd}{N(c+4nv)(d+4n\tilde{v})}\,
.
\end{equation}
For $N=1$, the ALE space becomes a flat $\mathbb{R}^4$, and we
obtain the result given in \cite{rg_int}. As in the previous
subsection, the solution describes an RG flow from a (2,0) CFT to a
(4,0) CFT in the IR. The central charges to curvature squared terms
are given by
\begin{eqnarray}
  \textrm{UV}: \, c_L&=&\frac{48\pi^2(c+4nv)(d+4n\tilde{v})N}{G^{(6)}_N}\left(1+
  \frac{\beta}{N\sqrt{(c+4nv)(d+4n\tilde{v})}}\right),\nonumber \\
  c_R&=&\frac{48\pi^2
  (c+4nv)(d+4n\tilde{v})N}{G_N^{(6)}},\\
  \textrm{IR}: \, c_L&=&\frac{48\pi^2cd}{G^{(6)}_N}\left(1+\frac{\beta}{cd}\right),\qquad c_R=\frac{48\pi^2
  cd}{G_N^{(6)}}\, .
\end{eqnarray}
\indent As in the previous subsection, it can be shown that this is
also a vev flow driven by a vev of a marginal operator of dimension
two.
\section{RG flows in type IIB and type I$'$ theories}\label{4Dflow}
In this section, we study an RG flow solution in type IIB theory on
an ALE background. Since there is no gauge field in type IIB theory,
the corresponding flow solution only involves gravitational
instantons. We also consider a solution in type I$'$ theory which is
a T-dual of the usual type I theory on $T^2$ and can also be
obtained from type IIB theory on $T^2/(-1)^{F_L}\Omega I_2$. As we
will see, in type I$'$ theory, there are more possibilities of the
gauge groups for the quiver gauge theory in the UV and, as a result,
more possible RG flows.
\subsection{RG flows in type IIB theory}\label{IIBflow}
We now study a supersymmetric flow solution in type IIB theory. We
begin with supersymmetry transformations of the gravitino $\psi_M$
and the dilatino $\chi$. These can be found in various places, see
for example \cite{IIB_schwarz, Howe_West}, and are given by
\begin{eqnarray}
\delta \chi &=&iP_M \Gamma^M
\epsilon^*-\frac{i}{24}F_{M_1M_2M_3}\Gamma^{M_1M_2M_3}\epsilon,
\nonumber \\
\delta \psi_M &=&\nabla_M \epsilon
-\frac{i}{1920}F^{(5)}_{M_1M_2M_3M_4M_5}\Gamma^{M_1M_2M_3M_4M_5}\Gamma_M
\epsilon \nonumber \\ & & +
\frac{1}{96}F_{M_1M_2M_3}(\Gamma_M^{\phantom{as}M_1M_2M_3}-9\delta^{M_1}_M\Gamma^{M_2M_3})\epsilon
\end{eqnarray}
where
\begin{eqnarray}
P_M &=&\frac{1}{2}(\pd_M\phi+ie^\phi\pd_M C_0),\nonumber \\
F_{M_1M_2M_3}&=&e^{-\frac{\phi}{2}}H_{M_1M_2M_3}+ie^{\frac{\phi}{2}}F_{M_1M_2M_3}\,
.
\end{eqnarray}
\indent In our ansatz, we choose $\phi=0$, $C_0=0$ and
$F_{M_1M_2M_3}=0$, so $\delta \chi=0$ is automatically satisfied.
The ten dimensional metric is given by
\begin{equation}
ds^2=e^{2f}dx^2_{1,3}+e^{2g}ds^2_4+e^{2h}(dr^2+r^2d\theta^2).\label{10Dmetric}
\end{equation}
The metric $ds^2_4$ is the ALE metric in \eqref{ALEmetric}, and the
functions $f$, $g$ and $h$ depend only on ALE coordinates $y^a$ and
$r$. We will use indices $\mu,\nu=0,\ldots, 3$, $a,b=4,\ldots, 8$.
The ansatz for the self-dual five-form field strength is
\begin{equation}
F^{(5)}=\tilde{F}+\hat{*}\tilde{F}\label{F5_ansatz}
\end{equation}
where $\hat{*}$ is the ten dimensional Hodge duality. We choose
$\tilde{F}$ to be
\begin{eqnarray}
\tilde{F}&=&dx^0\wedge dx^1\wedge dx^2\wedge dx^3\wedge
(U^{(1)}+Kdr)+rG^{(3)}\wedge dr\wedge d\theta
+r\tilde{G}^{(4)}\wedge d\theta, \nonumber \\
\hat{*}\tilde{F}&=&e^{-4f}(e^{2(g+h)}*U^{(1)}+e^{4g}*Krd\theta)+e^{4f-2(g+h)}dx^0\wedge
dx^1\wedge dx^2\wedge dx^3\wedge
*G^{(3)}\nonumber \\ & &+e^{4(f-g)}*\tilde{G}^{(4)} dx^0\wedge dx^1\wedge
dx^2\wedge dx^3\wedge dr
\end{eqnarray}
with all functions depending only on $y^a$ and $r$. We have used the
convention $\epsilon_{01234567r\theta}=1$. The notation $X^{(n)}$
means the $n$-form $X$ on the four dimensional space whose metric is
$ds^2_4$. The Bianchi identity $DF^{(5)}=0$ and self duality
condition impose the conditions
\begin{eqnarray}
dU^{(1)}&=&0,\qquad dK=\pd_r U^{(1)}\Rightarrow U^{(1)}=d\Lambda, \qquad K=\pd_r\Lambda+c_1, \nonumber \\
*G^{(3)}&=&e^{-4f+2(g+h)}d\Lambda,\qquad
*\tilde{G}^{(4)}=e^{-4(f-g)}K\, .
\end{eqnarray}
The $*$ and $d$ are the Hodge dual and exterior derivative on
$ds^2_4$, and $c_1$ is a constant.
\\
\indent From \eqref{10Dmetric}, we can read off the vielbein
components
\begin{equation}
e^{\hat{\mu}}=e^fdx^\mu,\qquad
e^{\hat{a}}=e^g\bar{e}^{\hat{a}},\qquad e^{\hat{r}}=e^hdr,\qquad
e^{\hat{\theta}}=e^hrd\theta\, .
\end{equation}
The $\bar{e}^{\hat{a}}$ is the vielbein on the ALE space. The spin
connections are given by
\begin{eqnarray}
\omega^{\hat{\theta}}_{\phantom{s}\hat{r}}&=&e^{-h}\left(\frac{1}{r}+h'\right)e^{\hat{\theta}},
\qquad \omega^{\hat{\theta}}_{\phantom{s}\hat{a}}=e^{-g}\pd_a
he^{\hat{\theta}},\qquad
\omega^{\hat{r}}_{\phantom{s}\hat{a}}=e^{-g}\pd_ahe^{\hat{r}}-e^{-h}g'e^{\hat{a}},\nonumber
\\
\omega^{\hat{a}}_{\phantom{s}\hat{b}}&=&e^{-g}(\pd_bg\delta^a_c-\pd^ag\delta^b_c)e^{\hat{c}}+e^{-g}\bar{\omega}^{\hat{a}}_{\phantom{s}\hat{b}},
\qquad
\omega^{\hat{\mu}}_{\phantom{s}\hat{a}}=e^{-g}\pd_afe^{\hat{\mu}},
\nonumber \\
\omega^{\hat{\mu}}_{\phantom{s}\hat{r}}&=&e^{-h}f'e^{\hat{\mu}}
\end{eqnarray}
where $\bar{\omega}^{\hat{a}}_{\phantom{s}\hat{b}}$ are spin
connections on the ALE space. We also use the following ten
dimensional gamma matrices
\begin{eqnarray}
\Gamma_{\hat{\mu}} &=&\gamma_{\hat{\mu}}\otimes \mathbf{I}_4\otimes
\mathbf{I}_2,\qquad \Gamma_{\hat{a}} =\tilde{\gamma}_5\otimes
\gamma_{\hat{a}}\otimes \mathbf{I}_2, \nonumber \\
\Gamma_{\hat{r}} &=&\tilde{\gamma}_5\otimes \hat{\gamma}_5\otimes
\sigma_1, \qquad \Gamma_{\hat{\theta}} =\tilde{\gamma}_5\otimes
\hat{\gamma}_5\otimes \sigma_2\, .
\end{eqnarray}
Throughout this paper, we use the notation $\mathbf{I}_n$ for
$n\times n$ identity matrix. The chirality condition on $\epsilon$
is $\Gamma_{11}\epsilon=\tilde{\gamma}_5\otimes
\hat{\gamma}_5\otimes \sigma_3\epsilon=\epsilon$.
$\tilde{\gamma}_5=i\gamma_0\gamma_1\gamma_2\gamma_3$ and
$\hat{\gamma}_5=\gamma_4\gamma_5\gamma_6\gamma_7$ are chirality
matrices in $x^\mu$ and $y^a$ spaces, respectively. With only 5-form
turned on, the relevant BPS equations come from
\begin{equation}
\delta \psi_M=\nabla_M\epsilon-\frac{i}{1920}F\Slash
^{(5)}\Gamma_M\epsilon
\end{equation}
where $ F\Slash
^{(5)}=F^{(5)}_{M_1M_2M_3M_4M_5}\Gamma^{M_1M_2M_3M_4M_5}$. It is now
straightforward to show that all the BPS equations are satisfied
provided that we choose
\begin{equation}
h=g=-f,\qquad \Lambda=2e^{4f},\qquad
\epsilon=e^{\frac{1}{2}f+\frac{i}{2}\sigma_3\theta}\hat{\epsilon}
\end{equation}
with $\hat{\epsilon}$ being the Killing spinor on the ALE space and
satisfying the condition
\begin{equation}
\bar{\nabla}_a\hat{\epsilon}=0\, .
\end{equation}
Furthermore, $\hat{\epsilon}$ satisfies a projection condition
$\gamma_5\hat{\epsilon}=\hat{\epsilon}$. So, the solution is again
$\frac{1}{4}$ supersymmetric along the flow. With these conditions
inserted in \eqref{F5_ansatz}, we obtain the 5-form field
\begin{equation}
F^{(5)}=2dx^0\wedge dx^1\wedge dx^2\wedge dx^3\wedge
\hat{d}\Lambda+2e^{-8f}\hat{*}\hat{d}\Lambda
\end{equation}
where now $\hat{*}$ and $\hat{d}$ are those on the six dimensional
space $\textrm{ALE}\times \mathbb{R}^2$ with coordinates
$(y^a,r,\theta)$. Equation $DF^{(5)}=0$ then gives
\begin{equation}
\hat{d}(e^{-8f}\hat{*}\hat{d}\Lambda)=\hat{d}\hat{*}\hat{d}e^{-4f}=0\,
.\label{harmonic_eq}
\end{equation}
So, the function $e^{-4f}$ satisfies a harmonic equation on
ALE$\times \mathbb{R}^2$.
\\
\indent It turns out to be difficult to find the explicit form of
this harmonic function. This function can be constructed from the
Green's function whose existence has been shown in \cite{existence},
see also \cite{etesi_existence}. We now consider the behavior of
this funciton at the two fixed points. The six dimensional metric is
given by
\begin{equation}
d\tilde{s}^2=V^{-1}(d\tau+\vec{\omega}.d\vec{x})^2+Vd\vec{x}.d\vec{x}+dr^2+r^2d\theta^2\,
.
\end{equation}
As $|\vec{x}|\rightarrow \infty$, with the coordinate changing given
in the previous section, the ALE metric become
$\mathbb{R}^4/\mathbb{Z}_N$. So, the metric of the whole six
dimensional space can be written as
\begin{equation}
d\tilde{s}^2=dR^2+R^2ds^2(S^5/\mathbb{Z}_N)
\end{equation}
where $R^2=4N|\vec{x}|+r^2$. \\ \indent Similarly, we can show that
as $\vec{x}\rightarrow \vec{x}_1$, the metric becomes the flat
$\mathbb{R}^6$ metric
\begin{equation}
d\tilde{s}^2=d\tilde{R}^2+\tilde{R}^2d\Omega_5^2
\end{equation}
where $\tilde{R}^2=4|\vec{x}-\vec{x}_1|+r^2$.
\\
\indent So, in order to interpolate between two conformal fixed
points, this function must satisfy the boundary condition
\begin{equation}
e^{-4f}\sim \frac{1}{R^4}\, .\label{boundary_condition}
\end{equation}
at both ends. There is also a relative factor of $N$ between the two
end points. This is due to the fact that the integral of the
harmonic equation \eqref{harmonic_eq} on $d\tilde{s}^2$ must vanish,
and this integral is in turn reduced to the integral of the gradient
of the Green's function over $S^5$ and $S^5/\mathbb{Z}_N$ at the two
end points. So, with all these requirements, the required harmonic
function has boundary conditions
\begin{eqnarray}
\vec{x}\rightarrow \vec{x}_1&:&\qquad
e^{-4f}=\frac{C}{R^4},\nonumber \\
|\vec{x}|\rightarrow \infty&:&\qquad e^{-4f}=\frac{CN}{R^4}\, .
\end{eqnarray}
The full metrics at both end points take the form
\begin{eqnarray}
\vec{x}\rightarrow \vec{x}_1&:&\qquad ds^2_{10}=\frac{R^2}{\sqrt{C}}dx^2_{1,3}+\frac{\sqrt{C}}{R^2}dR^2+\sqrt{C}d\Omega^2_5 \nonumber \\
|\vec{x}|\rightarrow \infty &:&\qquad
ds^2_{10}=\frac{\tilde{R}^2}{\sqrt{NC}}dx^2_{1,3}+\frac{\sqrt{NC}}{\tilde{R}^2}d\tilde{R}^2+\sqrt{NC}ds^2(S^5/\mathbb{Z}_N).
\end{eqnarray}
We obtain the two $AdS_5$ radii $L_1=C^{\frac{1}{4}}$ and $L_\infty
=(CN)^{\frac{1}{4}}$. The central charge is given by \cite{4D_c_a}
\begin{equation}
a=c=\frac{\pi L^3}{8G_N^{(5)}}\, .
\end{equation}
The ratio of the central charges is given by
\begin{equation}
\frac{a_1}{a_\infty}=\frac{c_1}{c_\infty}=\frac{L_1^8\textrm{Vol}(S^5)}{L_\infty^8\textrm{Vol}(S^5/\mathbb{Z}_N)}
=N\left(\frac{L_1}{L_\infty}\right)^8=\frac{1}{N}.
\end{equation}
The flow describes the deformation of $N=2$ quiver $SU(n)^N$ gauge
theory in the UV to $N=4$ $SU(n)$ SYM in the IR in which the gauge
group $SU(n)$ is the diagonal subgroup of $SU(n)^N$.
\\
\indent We now compute the central charges to curvature squared
terms. Higher derivative corrections to the central charges in four
dimensional CFTs have been considered in many references, see for
example \cite{4D_c_a, narain, Anselmi}. The five dimensional gravity
Lagrangian with higher derivative terms can be written as
\begin{equation}
\mathcal{L}=\frac{\sqrt{-g}}{2\kappa_5^2}(R+\Lambda+\alpha R^2+\beta
R_{\mu\nu}R^{\mu\nu}+\gamma
R_{\mu\nu\rho\sigma}R^{\mu\nu\rho\sigma}).
\end{equation}
$\Lambda$ is the cosmological constant. The central charges $a$ and
$c$ appear in the trace anomaly
\begin{eqnarray}
\langle T^\mu_{\phantom{a}\mu}\rangle
&=&\frac{c}{16\pi^2}\left(R_{\mu\nu\rho\sigma}R^{\mu\nu\rho\sigma}-2R_{\mu\nu}R^{\mu\nu}+\frac{1}{3}
R^2\right)\nonumber
\\ & &-\frac{a}{16\pi^2}(R_{\mu\nu\rho\sigma}R^{\mu\nu\rho\sigma}-4R_{\mu\nu}R^{\mu\nu}+
R^2).
\end{eqnarray}
Compare this with the holographic Weyl anomaly gives
\begin{eqnarray}
a&=&\frac{\pi
L^3}{8G^{(5)}_N}\left[1-\frac{4}{L^4}(10\hat{\alpha}+2\hat{\beta}+\hat{\gamma})\right],\nonumber
\\
c&=&\frac{\pi
L^3}{8G^{(5)}_N}\left[1-\frac{4}{L^4}(10\hat{\alpha}+2\hat{\beta}-\hat{\gamma})\right]
\end{eqnarray}
where we have separated the $AdS_5$ radius out of
$\alpha=\frac{\hat{\alpha}}{L^2}$, $\beta=\frac{\hat{\beta}}{L^2}$
and $\gamma=\frac{\hat{\gamma}}{L^2}$. Only $\gamma$ can be
determined from string theory calculation. Furthermore, there is an
ambiguity in $\alpha$ and $\beta$ due to field redefinitions. \\
\indent For $N=4$ SYM with gauge group $SU(n)$ in the IR, there is
no correction from $R^{\mu\nu\rho\sigma}R_{\mu\nu\rho\sigma}$ term.
To this order, the central charges are then given by
\begin{equation}
a_{IR}=c_{IR}=\frac{\pi^4
L^8}{8G_N^{(10)}}=\frac{\pi^4C^2}{8G_N^{(10)}}\, .
\end{equation}
\indent On the other hand, in the UV, we have $N=2$, $SU(n)^N$
quiver gauge theory. The central charges are
\begin{eqnarray}
a_{UV}&=&\frac{\pi^4
NC^2}{8G_N^{(10)}}\left[1-\frac{4}{NC}(10\hat{\alpha}+2\hat{\beta}+\hat{\gamma})\right],\nonumber
\\ c_{UV}&=&\frac{\pi^4
NC^2}{8G_N^{(10)}}\left[1-\frac{4}{NC}(10\hat{\alpha}+2\hat{\beta}-\hat{\gamma})\right].
\end{eqnarray}
The constant $C$ in our solution is related to the number of
D3-branes, $N_3$. The leading term in $a$ and $c$ is of order $C^2$
while the subleading one is of order $C$ as expected. The analysis
of the metric fluctuation can be carried out as in the six
dimensional case and gives $\Delta=2$. The flow is a vev flow driven
by a vacuum expectation value of a relevant operator of dimension
two.
\\
\indent Before discussing the RG flow on the dual field theory, let
us recall that ALE gravitational instantons admit a hyperkahler
quotient construction, which can be understood nicely  in terms of
the moduli space of a transverse (regular) D-brane probe moving off
the orbifold fixed point in $\mathbb{R}^4/\mathbb{ Z}_N$
\cite{douglas_quiver}, \cite{pol_K3Orientifold}. Starting with
$U(N)$ valued fields $X$, $\bar X$ on which one performs the
$\mathbb{ Z}_N$ projection, one denotes the invariant
(one-dimensional) components by $X_{i,i+1}$, $\bar{X}_{i+1,i}$, for
$i=0,\dots, N-2$, $X_{N-1,0}$, $\bar{X}_{0,N-1}$, which are the
links of the quiver diagram corresponding to the $A_{N-1}$ extended
Dynkin diagram. The resulting gauge group is $U(1)^N$, with a
trivially acting center of mass $U(1)$. It is convenient to
introduce the doublet fields $\Phi_r$
\begin{equation}
\Phi_r=\left(
    \begin{array}{c}
      X_{r-1,r} \\
      \bar{X}^{\dagger}_{r,r-1} \\
    \end{array}
  \right)
\end{equation}
for $r=1,\dots,N-1$, and

\begin{equation}
\Phi_0=\left(
    \begin{array}{c}
      X_{N-1,0} \\
      \bar{X}^{\dagger}_{0,N-1} \\
    \end{array}
  \right)
\end{equation}
After removing the trivial center of mass $U(1)$, the gauge group is
$U(1)^{N-1}$, and the $\Phi$'s have definite charges with respect to
it. After introducing Fayet-Iliopoulos (FI) terms $\vec{D}_r$,
$r=0,\dots, N-1$, with $\sum_r\vec{D}_r=0$, corresponding to closed
string, blowing-up mouduli, one gets the following potential:
\begin{equation}
U=
\sum_{r=0}^{N-1}\left(\Phi^\dagger_r\vec{\sigma}\Phi_r-\Phi^\dagger_{r+1}\vec{\sigma}\Phi_{r+1}+\vec{D}_r\right)^2.
\end{equation}
and the  $N-1$ independent D-flatness conditions, are then given by:
\begin{equation}
\Phi^\dagger_{r+1}\vec{\sigma}\Phi_{r+1}-\Phi^\dagger_r\vec{\sigma}\Phi_r=\vec{D}_r\,
\label{Dequation}.
\end{equation}
The ALE metric \eqref{ALEmetric} can be obtained after defining the
ALE coordinate and centers
\begin{equation}
\vec{x}=\Phi_0^\dagger\vec{\sigma}\Phi_0,\qquad
\vec{x}_i=\sum_{r=0}^{i-1}\vec{D}_r,\label{ALE_coordinate}
\end{equation}
respectively, and computing the gauge invariant kinetic term on the
$\Phi$'s, subject to the D-terms constraints
\cite{pol_K3Orientifold}.
\\
\indent This procedure can be generalized to the case of $n$ regular
D3-branes transverse to the ALE space. Starting with $U(nN)$ valued
Chan-Paton factors, the resulting theory after projection, is the
$N=2$  $SU(n)^N$ gauge theory, with hypermultiplets formed by the
fields $X_{ij}$ and $\bar{ X}_{ij}$ related to the links of the
quiver diagram as above, but now in the $(n,\bar{ n})$, $(\bar
{n},n)$ representations of the $SU(n)$'s at the vertices of the
quiver diagram. In addition, there are adjoint scalars $W_i$ in the
adjoint of $SU(n)$, belonging to the vector multiplets. The theory
is conformally invariant and describes the dual $N=2$ SCFT at the UV
point.
\\
\indent In order to match with  the RG flow from the UV to IR
described previously on the gravity side, which gives an $N=4$
theory in the IR, we consider the Higgs branch of the $N=2$ UV
theory discussed above. Therefore, we set $\langle W_i\rangle=0$ and
give vev's to the hypermultiplets $X_{ij}$, $\bar{X}_{ij}$. The
equations governing the vacua of the theory are then the obvious
matrix generalization of \eqref{Dequation}, with $N-1$ independent
triplets of FI terms for the $N-1$  $U(1)$'s, in an $SU(2)_R$
invariant formulation or can be written in and $N=1$ fashion
directly in terms of $X_{ij}$, $\bar{X}_{ij}$ and their hermitean
conjugates. In any case, it is clear that by giving digonal vev's to
$X$ 's ($\bar X$'s)
\begin{equation}
\langle X_{ij}\rangle=x_{ij} \mathbf{I}_n,\qquad \textrm{for all}\,
i,j
\end{equation}
compatible with the D-flatness conditions, we can break $SU(n)^N$
down to the diagonal $SU(n)$, with a massless spectrum coinciding
with that of $N=4$ SYM theory for $SU(n)$ gauge group. A similar
flow, from $N=1$ to $N=4$, has been studied in \cite{chethan} and
\cite{nessi} in the case of the ALE space $\mathbb{ C}_3/\mathbb{
Z}_3$.
\\
\indent Notice that we can have intermediate possibilities for the
IR point. In terms of the geometry, this can happen when some of the
ALE centers $x_i$ coincide with each other. Recalling the ALE
metric, we have already seen that in the UV
\begin{displaymath}
V\sim \frac{N}{|\vec{x}|},\qquad |\vec{x}|\rightarrow \infty\, .
\end{displaymath}
In the IR, if we let $M$ centers, $M<N$ to coincide with $\vec{x}_1$
say, and zoom near $\vec{x}_1$, we have
\begin{equation}
V\sim \frac{M}{|\vec{x}-\vec{x}_1|}, \qquad \vec{x}\rightarrow
\vec{x}_1\, .
\end{equation}
The ALE geometry then develops a $ \mathbb{ Z}_M$ singularity in the
IR. Therefore, all possibilities with any values of $N$ and $M$
should be allowed as long as $M<N$. We can also compute the ratio of
the central charges by repeating the same procedure as in the
previous section and end up with the result
\begin{equation}
\frac{a_{IR}}{a_{UV}}=\frac{c_{IR}}{c_{UV}}=\frac{M}{N}<1\, .
\end{equation}
On the other hand, on the field theory side, we can partially Higgs
the gauge group $SU(n)^N$ down to $SU(n)^M$, for any $M<N$. That is
we have flows between the corresponding quiver diagrams.
\subsection{RG flows in Type I$'$ string theory}
As we mentioned, type I$'$ is obtained from type I theory by two
T-duality transformations along the two cycles of $T^2$. In this
process, D9-branes will become D7-branes and the $SO(32)$ gauge
group is broken to $SO(8)^4$, corresponding to the four fixed points
of $T^2$. It has been shown in \cite{Sen_K3F_theory}, that the
resulting theory is dual to type IIB theory on $T^2/(-1)^{F_L}\Omega
I_2$. One then considers a stack of D3-branes near one of the fixed
points and in the near horizon geometry one gets $AdS_5\times
S^5/\mathbb{Z}_2$. This corresponds to a dual $N=2$ CFT, with
$USp(2n)$ gauge group and $SO(8)$ global flavor symmetry
\cite{spalinski, aharony}, with matter hypermultiplets in the
antisymmetric representation of $USp(2n)$ and also in the (real)
$(2n, 8)$ of $USp(2n)\times SO(8)$. In our case, we are replacing
$\mathbb{R}^4$ with ALE space, or in the orbifold limit, with
$\mathbb{R}^4/ \mathbb{Z}_N$. Similar to the type IIB case, the UV
field theory will be obtained by performing the orbifold
$\mathbb{Z}_N$ projection of the above field content, which in turn
will be recovered at the IR point after Higgsing.
\\
\indent On the supergravity side, we will restrict our analysis to
the two-derivative terms in the affective action. Therefore, the
ansatz of the previous subsection can be carried over to this case.
In particular, the Bianchi identity for the 5-form will be
unchanged. Otherwise, one would have to switch on also D7-brane
instantons on the ALE space in order to compensate for the $R\wedge
R$ term present on the right-hand side of the Bianchi identity at
order $\mathcal{O}(\alpha')$. The analysis in this case is closely
similar to the previous case apart from the facts that we start with
16 supercharges in ten dimensions rather than 32, and the final
equation for $e^{-4f}$ is the same as before. Following similar
analysis as in the previous subsection, we can show that the
solution interpolates between $AdS_5\times S^5/(\mathbb{Z}_N\times
\mathbb{Z}_2)$ in the UV and $AdS_5\times S^5/\mathbb{Z}_2$ in the
IR, with $\mathbb{Z}_2$ being the orientation reversal operator
$\Omega$.
\\
\indent As mentioned above, the field theory interpretation will
involve flows between $N=2$ quiver gauge theories with different
gauge groups. The gauge group in the UV will be obtained by
considering orbifolding/orientifolding a system of D3/D7 branes,
whereas the IR group will be obtained by Higgsing, like in type IIB
case. According to \cite{douglas_quiver}, the choice of gauge groups
depends on the values of $N$ as well as on the choice of a
$\mathbb{Z}_N$ phase relating $\Omega$ and $\mathbb{Z}_N$
projections. In our case, the D3-brane worldvolume gauge theory
descends from the theory on D5-brane for which
$\gamma(\Omega)^t=-\gamma(\Omega)$. In what follows, we will use the
notations of \cite{douglas_quiver} and also refer the reader to this
reference for more detail on the quiver gauge theory. We first
review the consistency conditions for the $\mathbb{Z}_N\times
\mathbb{Z}_2$ actions \cite{douglas_quiver}
\begin{eqnarray}
\Omega^2=1&:&\qquad
\gamma(\Omega)=\chi(\Omega)\gamma(\Omega)^t,\nonumber \\
\Omega g=g \Omega &:&\qquad
\gamma(g)\gamma(\Omega)\gamma(g)^t=\chi(g,\Omega)\gamma(\Omega),\nonumber
\\
g^n=1&:&\qquad \gamma(g)^n=\chi(g)1
\end{eqnarray}
where $g\in \mathbb{Z}_N$ and $\chi(\Omega)$, $\chi(g)$ and
$\chi(g,\Omega)$ are phases. As shown in \cite{douglas_quiver}, we
can set $\chi(g)=1$. Furthermore, we are interested in the case of
$\chi(\Omega)=-1$ on the D3-branes and $\chi(\Omega)=+1$ on the
D7's. In type I theory, there are five cases to consider, but only
three of them are relevant for us. These are $N$ odd,
$\chi(g,\Omega)=1$, $N$ even, $\chi(g,\Omega)=1$ and $N$ even,
$\chi(g,\Omega)=\xi$ with $\xi=e^{\frac{2\pi i}{N}}$. We
now consider RG flows in these cases. \\
\subsubsection{$\chi(g,\Omega)=1$, $N$ $\textrm{odd}$}
\begin{figure}[h]
  \centering
  \includegraphics[width=11cm]{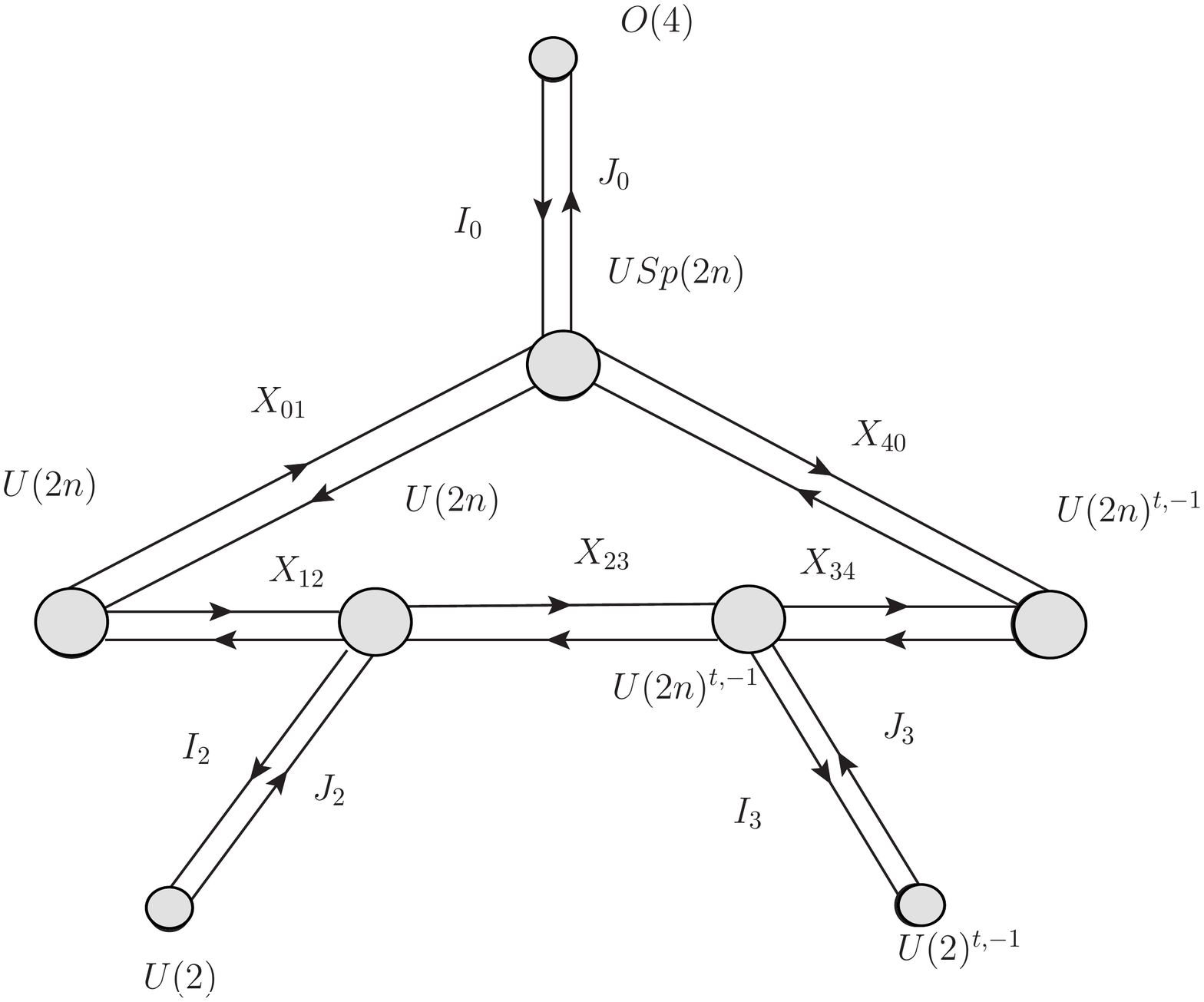}\\
  \caption{Quiver diagram for $\chi(\Omega)=-1$, $\chi(g,\Omega)=1$ and $N=5$.}
  \label{quiver1}
\end{figure}

In this case, the gauge group is given by
\begin{eqnarray}
G_1&=&USp(v_0)\times [U(v_1)\times U(v_2)\times \ldots \times
U(v_{\frac{N-1}{2}})]\nonumber \\
&=&\{U_0,U_1,\ldots , U_{N-1}|U_iU^t_{N-i}=1,1\leq i\leq
N-1\}.\label{gauge_group1}
\end{eqnarray}
Our convention is that $USp(2n)$ has rank $n$. The full
configuration involves also the quiver theory on D9-branes which
give rise to D7-branes in our case. Our main aim here is to study
the symmetry breaking of the gauge group on D3-branes. The presence
of D7-branes is necessary to make the whole system conformal. For
the UV quiver gauge theory to be conformal, we choose
$v_0=v_1=\ldots = v_{\frac{N-1}{2}}=n$ with an appropriate number of
D7-branes such that the field theory beta function vanishes. Using
the notation of \cite{douglas_quiver}, we denote the vector spaces
associated to the nodes of the inner quiver, the D3-branes, by $V_i$
and those of the outer quiver on D7-branes by $W_i$. There is also
an identification of the nodes $V_i=V_{N-i}$ and similarly for
$W_i$'s, see \cite{douglas_quiver}. This condition gives rise to the
relation between the gauge groups of different nodes as shown in
\eqref{gauge_group1}.
\\
\indent The gauge theory on the D7-branes is described by similar
gauge group structure but with $USp(2n)$ replaced by $O(2n)$ due to
the opposite sign of $\chi(\Omega)$. In addition to the vector
multiplets, there are hypermultiplets, $X$, $\bar X$, associated to
the links connecting the $V_i$'s and $I,J$ related to the links
connecting $V_i$'s and $W_i$'s, in bifundamental representations of
the respective gauge groups. The vanishing of the beta function can
be achieved by setting $w_0=4$,
$w_{\frac{N-1}{2}}=w_{\frac{N+1}{2}}=2$ and the other $w_i$'s zero.
Notice that the non trivial gauge groups on the outer quiver are
associated with two types of the nodes of the inner quiver. The
first type consists of the nodes with $USp$ gauge groups while the
second type contains nodes connected to each other by antisymmetric
scalars. The corresponding outer gauge groups for these two types
are $SO(4)$ and $U(2)$, respectively.
\\
\indent It is easy to see the reason for this pattern of inner/outer
gauge groups: the point is that for the inner nodes with $U(2n)$
gauge groups and connected by $U(2n)$ bifundamental scalars, the
corresponding part of the quiver diagram is essentially the same as
the quiver diagram arising from type IIB theory in which all the
gauge groups are unitary. It is well-known that this quiver gauge
theory is supeconformal without any extra field contents.
\\
\indent As observed in \cite{douglas_quiver, intriligator}, the
above construction matches with the ADHM construction of $SO(n)$
instantons on ALE spaces: for example, the assignement of D7-brane
gauge group given above means that, at the boundary of the ALE
space, which has fundamental group $\pi_1=\mathbb{Z}_N$, the $SO(8)$
flat connection has holonomy which breaks $SO(8)$ down to
$SO(4)\times U(2)\sim SO(4)\times SU(2)\times U(1)$. On the other
hand, $G_1$ is the ADHM gauge group, related to the number of
instantons (D3-branes).
\\
\indent We now consider a Higgsing of this theory, and we need to be
more precise about the representations of the matter fields. The
nodes are connected to each other by the bifundamental scalars $X$
and $\bar{X}$. These scalars are subject to some constraints given
by
\begin{eqnarray}
X_{01}&=&-(X_{N-1,0}\omega_{2n})^t,\qquad
X_{\frac{N-1}{2},\frac{N+1}{2}}=-(X_{\frac{N-1}{2},\frac{N+1}{2}})^t,\nonumber
\\
X_{i,i+1}&=&(X_{N-i-1,N-i})^t, \qquad 1\leq i\leq
\frac{N-3}{2},\nonumber \\
\bar{X}_{10}&=&(\omega_{2n}\bar{X}_{0,N-1})^t,\qquad
\bar{X}_{\frac{N+1}{2},\frac{N-1}{2}}=-(\bar{X}_{\frac{N+1}{2},\frac{N-1}{2}})^t,\nonumber
\\
\bar{X}_{i+1,i}&=&(\bar{X}_{N-i,N-i-1})^t,\qquad 1\leq i\leq
\frac{N-3}{2}
\end{eqnarray}
where $\omega_{2n}$ represents the symplectic form of dimension
$2n$. We will show that after the RG flow, the theory will flow to
$USp(2n)$, $N=2$ gauge theory in which the gauge group $USp(2n)$ is
the diagonal subgroup of the $USp(2n)$ and $USp(2n)$ subgroups of
all the $U(2n)$'s. We first illustrate this with the simple case of
$N=5$. The corresponding quiver diagram is shown in Figure
\ref{quiver1}. In the figure, the outer quiver and the inner one are
connected to each other by scalar fields $I_i,J_i$ (we have omitted
the $\bar{X}$'s on the diagram). Notice that the gauge groups in the
outer quiver are orthogonal and unitary groups due to the opposite
sign of $\chi(\Omega)$. We will be interested in the Higgs branch,
i.e. we set the vev's of the scalars in the vector multiplets to
zero. Furthermore, we will also set $\langle I\rangle=\langle
J\rangle=0$ in all the cases we will discuss in the following. The
D-flatness conditions are then obtained from those of the type IIB
case by suitable projections/identifications on the $X$'s and $\bar
X$'s. The main difference, compared to the type IIB case, comes from
the gauge group, which involve an $USp(2n)$ factor at the 0-th
vertex and has $U(2n)$ factors which are related in the way
indicated in Figure \ref{quiver1}: as a result, the corresponding
FI terms obey $\vec{D}_0=0$, $\vec{D}_1=-\vec{D}_4$,
$\vec{D}_2=-\vec{D}_3$, with similar relations for higher odd $N$.
\\
\indent We will give only the flows in which $X$ and $\bar{X}$
acquire vev's. The above conditions then give
\begin{equation}
X_{01}=-X^t_{40},\qquad X_{12}=X^t_{34},\qquad
X_{23}=-X_{23}^t\label{vev1}
\end{equation}
and similarly for $\bar{X}$. We choose the vev's as follows
\begin{equation}
\langle X_{01}\rangle=a \mathbf{I}_{2n},\qquad \langle
X_{12}\rangle=b\mathbf{I}_{2n},\qquad \langle X_{23}\rangle
=c\omega_{2n}
\end{equation}
where $a$, $b$ and $c$ are constants. The vev's for $\bar{X}$ are
similar but with different parameters $\bar{a}$, $\bar{b}$ and
$\bar{c}$. Notice also that we only need to give vev's to the
independent fields since the vev's of other fields can be obtained
from \eqref{vev1}. From now on, we will explicitly analyze only the
$X$'s. The analysis for $\bar{X}$'s follows immediately.
\\
\indent The field $X_{ij}$ transforms as $g_iX_{ij}g^{-1}_j$ where
$g_i$ and $g_j$ are elements of the two gauge groups, $G_i$ and
$G_j$, connected by $X_{ij}$. The unbroken gauge group is the
subgroup of $USp(2n)\times U(2n)\times \ldots \times U(2n)$ that
leaves all these vev's invariant. The invariance of $X_{01}$
requires that $g_1$ is a symplectic subgroup of $U(2n)$ and
$g_1=g_0$. The invariance of $X_{12}$ imposes the condition
$g_1=g_2=g_0$ and so on. In the end, we find that the gauge group in
the IR is $USp(2n)_{\textrm{diag}}$. For any odd $N$, the whole
process works in the same way apart from the fact that there are
more nodes similar to $X_{12}$. These nodes can be given vev's
proportional to the identity. Taking this into account, we end up
with scalar vev's
\begin{eqnarray}
\langle X_{01}\rangle &=&a_{01}\mathbf{I}_{2n},\qquad \langle
X_{\frac{N-1}{2},\frac{N+1}{2}}\rangle=a_{\frac{N-1}{2},\frac{N+1}{2}}\omega_{2n},\nonumber \\
\langle X_{i,i+1}\rangle &=&a_{i,i+1}\mathbf{I}_{2n}, \qquad 1\leq
i\leq \frac{N-3}{2},
\end{eqnarray}
and the unbroken gauge group is $USp(2n)_{\textrm{diag}}$. In
addition, one can verify that the masless spectrum is precisely that
of the superconformal $USp(2n)$ theory with $SO(8)$ global symmetry
described at the beginning of this section.
\subsubsection{$\chi(g,\Omega)=1$, $N$ $\textrm{even}$}
\begin{figure}[h]
  \centering
  \includegraphics[width=11cm]{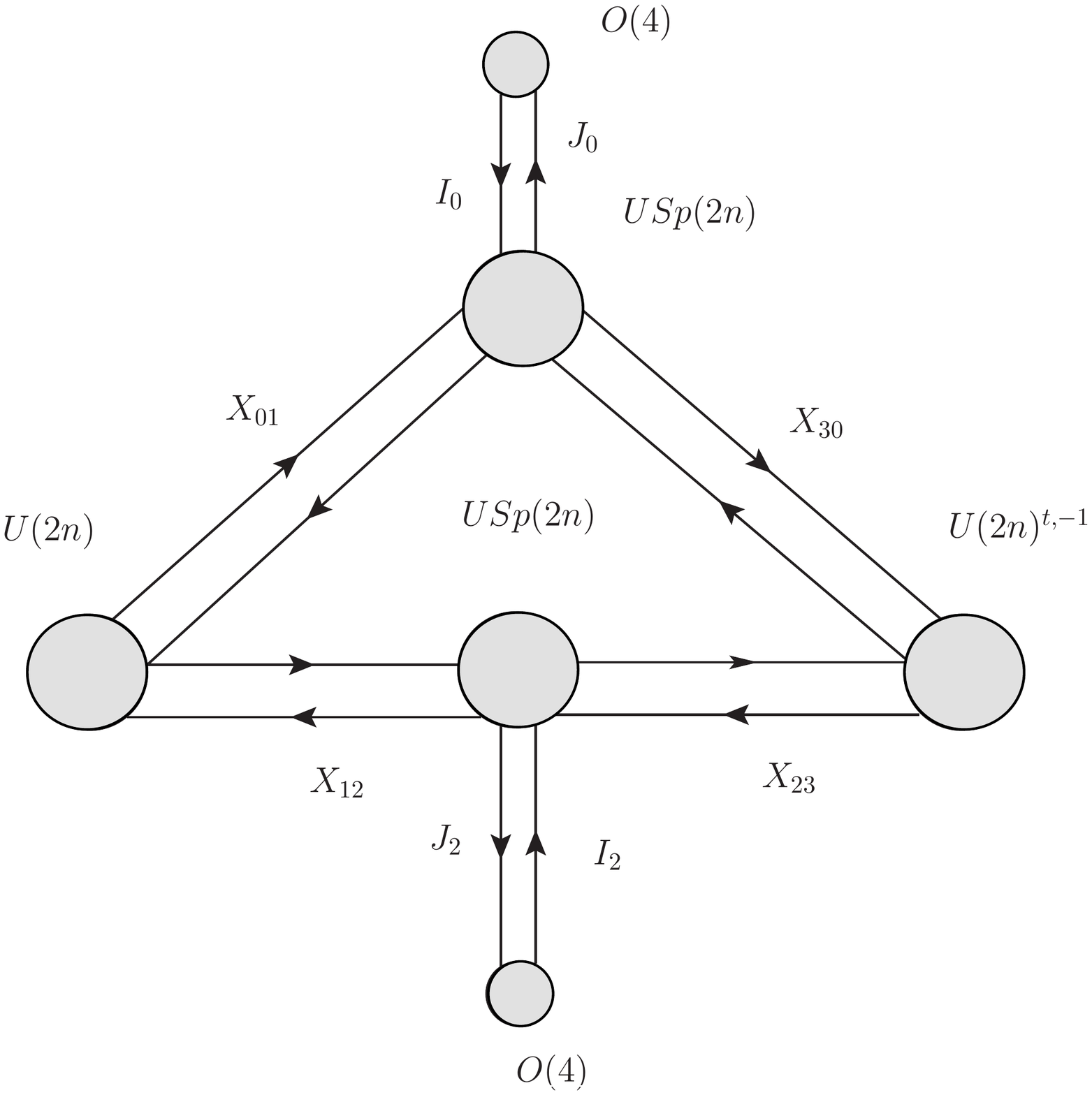}\\
  \caption{Quiver diagram for $\chi(\Omega)=-1$, $\chi(g,\Omega)=1$ and $N=4$.}
  \label{quiver2}
\end{figure}

In this case, we have the gauge group
\begin{eqnarray}
G_2&=&USp(v_0)\times [U(v_1)\times \ldots \times
U(v_{\frac{N}{2}-1})]\times USp(v_{\frac{N}{2}})\nonumber \\
&=&\left\{U_0,\ldots , U_{N-1}|U_iU^t_{N-i}=1, 1\leq i \leq
N-1,i\neq \frac{N}{2}\right\}.
\end{eqnarray}
Compared to the previous case, there is an additional
$USp(v_{\frac{N}{2}})$ gauge group at the
$\frac{N}{2}^{\textrm{th}}$ node. As before, we choose
$v_0=v_1=\ldots =v_{\frac{N}{2}}=2n$ and $w_0=w_{\frac{N}{2}}=4$
with other $w_i$'s being zero, corresponding to the breaking of the
D7-brane gauge group from $SO(8)$ down to $SO(4)\times SO(4)$. The
scalars are subject to the constraints
\begin{eqnarray}
X_{01}&=&\omega_{2n}(X_{N-1,o})^t,\qquad
X_{\frac{N}{2},\frac{N+2}{2}}=-\omega_{2n}(X_{\frac{N-2}{2},\frac{N}{2}})^t,\nonumber
\\
X_{i,i+1}&=&(X_{N-i-1,N-i})^t,\qquad 1\leq i< \frac{N-2}{2}\, .
\end{eqnarray}
The corresponding quiver diagram for $N=4$ is shown in Figure
\ref{quiver2}.
\\
\indent As for the FI terms in this case, clearly
$\vec{D}_0=\vec{D}_2=0$ and $\vec{D}_1=-\vec{D}_3$, with the obvious
generalization for higher even $N$. We can choose the following
vev's to Higgs the theory
\begin{eqnarray}
\langle X_{01}\rangle &=& x_{01}\mathbf{I}_{2n},\qquad \langle
X_{\frac{N-2}{2},\frac{N}{2}}\rangle
=x_{\frac{N-2}{2},\frac{N}{2}}\mathbf{I}_{2n},\nonumber
\\ \langle X_{i,i+1}\rangle &=&x_{i,i+1}\mathbf{I}_{2n},\qquad
1\leq i < \frac{N-2}{2}\, .
\end{eqnarray}
The symmetry breaking is the same as in the previous case. These
vev's are invariant under the unbroken gauge group
$USp(2n)_{\textrm{diag}}$, and one can verify that massless
hypermultiplets fill the spectrum of the $N=2$ theory discussed in
the previous case.
\subsubsection{$\chi(g,\Omega)=\xi$, $N$ $\textrm{even}$}
\begin{figure}[h]
  \centering
  \includegraphics[width=9cm]{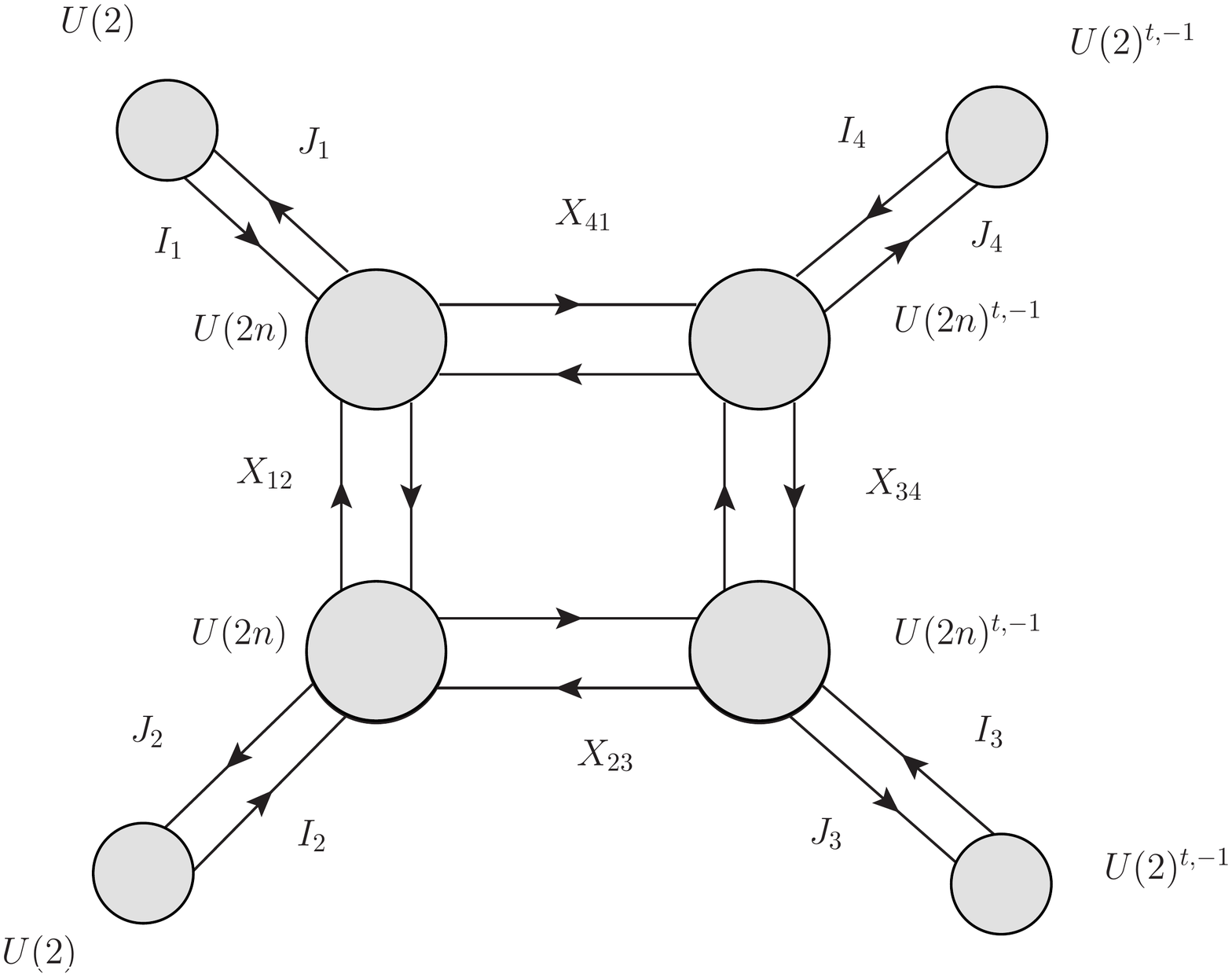}\\
  \caption{Quiver diagram for $\chi(\Omega)=-1$, $\chi(g,\Omega)=i$ and $N=4$.}
  \label{quiver3}
\end{figure}

It is possible to choose $\chi(g,\Omega)=\xi$ for $N$ even as shown
in \cite{douglas_quiver}, and this is our last case. We adopt the
range of the index $i$ from $1$ to $N$ in this case. The relevant
gauge group is given by
\begin{eqnarray}
G_3&=&U(v_1)\times U(v_2)\times \ldots \times
U(v_{\frac{N}{2}})\nonumber \\
&=&\{U_1,\ldots , U_N|U_iU^t_{N-i+1}=1,1\leq i\leq N\}.
\end{eqnarray}
We are interested in the case $v_1=v_2=\ldots =v_{\frac{N}{2}}=2n$
and $w_1=w_{\frac{N}{2}}=2$ with other $w_i$'s being zero, i.e. the
D7 gauge group is now broken down to $U(2)\times U(2)$. The
conditions on the scalar fields are
\begin{eqnarray}
X_{N1}&=&-X^t_{N1},\qquad X_{\frac{N}{2},\frac{N+2}{2}}=-(X_{\frac{N}{2},\frac{N+2}{2}})^t,\nonumber \\
X_{i,i+1}&=&(X_{N-i,N-i+1})^t,\qquad 1\leq i\leq \frac{N-2}{2}\, .
\end{eqnarray}
The quiver diagram for $N=4$ and $\xi=i$ is shown in Figure
\ref{quiver3}. Notice the relations $\vec{D}_1=-\vec{D}_4$,
$\vec{D}_2=-\vec{D}_3$ and so on for higher even $N$. There are two
possibilities for Higgsing this theory. The first one involves only
the vev's
\begin{equation}
\langle X_{i,i+1}\rangle =b_{i,i+1}\mathbf{I}_{2n},\qquad 1\leq
i\leq \frac{N-2}{2}\, .
\end{equation}
The unbroken gauge group is the diagonal subgroup of $U(v_1)\times
\ldots \times U(v_{\frac{N}{2}})$, $U(2n)_{\textrm{diag}}$. The
second possibility is to give vev's to all scalars including the
antisymmetric ones
\begin{eqnarray}
\langle X_{i,i+1}\rangle &=&b_{i,i+1}\mathbf{I}_{2n},\qquad 1\leq
i\leq \frac{N-2}{2}, \nonumber \\
\langle X_{N1}\rangle &=& b_{N1}\omega_{2n},\qquad \langle
X_{\frac{N}{2},\frac{N+2}{2}}\rangle
=b_{\frac{N}{2},\frac{N+2}{2}}\omega_{2n}\, .
\end{eqnarray}
In this case, the resulting gauge group is further broken down to
$USp(2n)_{\textrm{diag}}$.
\section{Symmetry breaking and geometric
interpretations}\label{geometry} In this section, like in the type
IIB case, we consider more general symmetry breaking patterns in the
field theory and match them with the  possible flows emerging from
the supergravity solution. This involves the cases in which the
gauge groups in the quiver gauge theory are not completely broken
down to a single diagonal subgroup. After symmetry breaking, the IR
CFT is again a quiver gauge theory with a reduced number of gauge
groups, and of course, the number of nodes is smaller. We will show
that some symmetry breaking patterns are not possible on the field
theory side, at least by giving simple vev's to scalar fields.
\\
\indent We now consider the possibility of RG flows from a UV CFT
which is a quiver gauge theory with the corresponding geometry
$AdS_5\times S^5/(\mathbb{Z}_N\times \mathbb{Z}_2)$ to an IR CFT
which is associated to the geometry $AdS_5\times
S^5/(\mathbb{Z}_M\times \mathbb{Z}_2)$ and $M<N$. We saw that in the
type IIB case this was always possible, and geometrically it was
related to geometries developing a $\mathbb{Z}_M$ orbifold
singularity obtained by bringing $M$ centers together in the smooth
ALE metric.
\\
\indent Let us start from the field theory side. It is easy to see
that it is not always possible to have a flow from one quiver
diagram to the other. For example, we consider a flow from the
diagram in Figure \ref{quiver1}, $N=5$, to Figure \ref{quiver2},
$N=4$. This can be done by giving a vev to $X_{23}$ and
$\bar{X}_{32}$ which transform in the antisymmetric tensor
representation of $U(2n)$. The gauge group $U(2n)$ at the node $v_2$
and $v_3$ will be broken to $USp(2n)$. The resulting IR theory is
then described by the quiver diagram in Figure \ref{quiver2}.
Continuing the process by Higgsing Figure \ref{quiver2} to the
diagram with $N=3$, we find that it is not possible to completely
break the $USp(2n)$ gauge group at $v_2$ with the remaining scalars
transforming in the antisymmetric tensor representation of $U(2n)$.
It might be achieved by giving a vev to complicated composite
operators, but we have not found any of these operators. Note also
that this is the case only for reducing the value of $N$ by one
unit. If we Higgs the $N=6$ to $N=4$ or in general $N$ to $N-2$,
this flow can always be achieved by giving vev's to $X_{12}$ and
$\bar{X}_{21}$. The gauge groups $U(2n)$ at $v_1$ and $v_2$ as well
as at their images $v_{N-1}$ and $v_{N-2}$ will be broken to
$U(2n)_{\textrm{diag}}$. The resulting quiver diagram is the same
type as the original one with two nodes lower. What we are
interested in is the problematic cases in which the flow connects
two types of diagrams and lowers $N$ by one unit.
\\
\indent We now begin with a diagram of the type shown in Figure
\ref{quiver3}. As mentioned in the previous section, this type is
only possible for even $N$. It is easily seen that giving a vev to
the $U(2n)$ antisymmetric scalars $X_{1N}$ and $\bar{X}_{N1}$
reduces the diagram to the $N-1$ diagram of the type shown in Figure
\ref{quiver1}. Furthermore, a diagram with $N-2$ nodes of the type
in Figure \ref{quiver2} can be obtained by giving an additional vev
to $X_{\frac{N}{2},\frac{N+2}{2}}$ and
$\bar{X}_{\frac{N+2}{2},\frac{N}{2}}$. Now, the problem arises in
deriving this diagram from the odd $N$ diagram. As before, the
$USp(2n)$ gauge groups at $v_0$ must be completely broken leaving
only scalars in the $U(2n)$ antisymmetric tensor. Actually, it seems
to be impossible to obtain this type of quiver diagrams from any of
the other two types by Higgsing in a single or multiple steps since
the process involves the disappearance of the $USp$ gauge group.
\\
\indent We now discuss how the above field theory facts match with
the geometry on the supergravity side. We will follow the approach
in \cite{pol_K3Orientifold}, where some peculiarities of type I
string theory on $\mathbb{ Z}_N$ orientifolds where clarified. The
idea is to use a regular D1-brane (in type I theory), to probe the
background geometry, following the same logic  explained for the
type IIB case in the previous subsection. In that case, we saw that
one could reproduce the full smooth ALE geometry by switching on FI
terms, which are background values of closed string moduli. The
$\mathbb{ Z}_N$ projection has generically the effect of reducing
unitary groups down to $SO/USp$ subgroups and/or of identifying
pairs of unitary groups, in a way which depends on the details of
the projection. We can indeed consider a probe D1-brane in the
present orientifold context and derive its effective field theory
for the three cases discussed in the previous section by assuming an
orthogonal projection $\chi(\Omega)=1$. In the following, the
diagrams in Figures \ref{quiver1}, \ref{quiver2} and \ref{quiver3}
will be referred to as type I, II and III quivers, respectively.
\\
\indent For the case 1, $\chi(g,\Omega)=1$, $N=2m+1$ odd, we will
have $m$ pairs of conjugate $U(1)$'s as gauge groups, (with an
$O(1)$ ``gauge group'' at the 0-th vertex of the inner quiver
diagram of Figure \ref{quiver1}), with appropriate identifications
of the scalars $X$, $\bar X$. For example for $N=3$, we have
$X_{01}=X_{20}$ plus $X_{12}$, similarly for $\bar X$ fields.
Consequently, for the FI terms, we will have $D_0=0$ and
$D_i=-D_{N-i}$, $i=1,\dots, m$. Translating these data to the ALE
centers $\vec{x}_i$ via \eqref{ALE_coordinate} as in the previous
subsection, we see that for $N=2m+1$ there are $m$ $\mathbb{ Z}_2$
singularities. There is in addition a simple pole in the function
$V$, which is however a smooth point in the geometry as long as it
is kept distinct from the other poles. The function $V$ in the ALE
metric \eqref{ALEmetric} is then given by
\begin{equation}
V=\frac{1}{|\vec{x}-\vec{x}_1|}+\sum_{i=2}^{m+1}\frac{2}{|\vec{x}-\vec{x}_i|}\,
.\label{V1}
\end{equation}
\indent If we choose the IR point by setting $\vec{x}\rightarrow
\vec{x}_1$, we end up with the flow from $N=2$ quiver gauge theory
of type I to the $N=2$, $USp(2n)_{\textrm{diag}}$ gauge theory. The
flow from type I quiver with $N=2m+1$ to type I quiver with $N=2m-1$
can be obtained by choosing $\vec{x}\rightarrow \vec{x}_1$ with
$\vec{x}_i=\vec{x}_1$ for $i=2,\ldots, m-1$. Finally, the flow to
type II quiver in the case 2 can be achieved by setting
$\vec{x}_i=\vec{x}_2$ for $i=3,\ldots, m-1$ and $\vec{x}\rightarrow
\vec{x}_2$.
\\
\indent For the case 2, $\chi(g,\Omega)=1$, $N=2m$ even, we will
have $O(1)$ at the nodes 0 and $m$, and the remaining $U(1)$'s are
pairwise conjugate, and there are obvious identifications for the
$X$ and $\bar X$ fields. Consequently, $\vec{D}_0=\vec{D}_m=0$ and
$\vec{D}_i=-\vec{D}_{N-i}$, $i=1,\dots,m-1$. In terms of the ALE
metric, we see that there are $m$ $\mathbb{ Z}_2$ singularities. The
corresponding $V$ function is
\begin{equation}
V=\sum_{i=1}^m\frac{2}{|\vec{x}-\vec{x}_i|}\, . \label{V2}
\end{equation}
\indent The possible flows are the following. First of all, to
obtain the $N=2$, $USp(2n)_{\textrm{diag}}$ gauge theory in the IR,
we choose $\vec{x}\rightarrow \vec{x}'$ where $\vec{x}'$ is any
regular point. The full Green function $G(x,x')$ will behave in the
same way as $\vec{x}\sim \vec{x}_i$. In this case, the IR geometry
is a smooth space. Another possible flows are given by Higgsing type
II diagram with $N=2m$ to the same type with $N=2m-2$. This is
achieved by setting $\vec{x}_i=\vec{x}_1$ for $i=2,\ldots, m-1$ and
$\vec{x}\rightarrow \vec{x}_1$.
\\
\indent Finally, for the case 3, $\chi(g,\Omega)=\xi$, therefore
$N=2m$ even, we have $m$ pairs of conjugate $U(1)$ factors and
consequently $\vec{D}_i=-\vec{D}_{N-1-i}$, $i=0,\dots,m$, which
implies $m-1$ $\mathbb{ Z}_2$ singularities plus two smooth points
in the geometry. The function $V$ is given by
\begin{equation}
V=\frac{1}{|\vec{x}-\vec{x_1}|}+\frac{1}{|\vec{x}-\vec{x}_{2m}|}+\sum_{i=2}^{m}\frac{2}{|\vec{x}-\vec{x}_i|}\,
.\label{V3}
\end{equation}
\indent The flow from type III quiver to $N=2$,
$USp(2n)_{\textrm{diag}}$ gauge theory is given by
$\vec{x}\rightarrow \vec{x}_1$ or $\vec{x}\rightarrow \vec{x}_{2m}$.
If we choose $\vec{x}\rightarrow \vec{x}_1$ and
$\vec{x}_i=\vec{x}_1$ for $i=2,\ldots, 2m-1$, we obtain the flow
from type III quiver with $N=2m$ to type I quiver with $N=2m-1$. On
the other hand, if we choose $\vec{x}\rightarrow \vec{x}_2$ and
$\vec{x}_i=\vec{x}_2$ for $i=3,\ldots, 2m-1$, we find a flow from
type III quiver with $N=2m$ to type II quiver with $N=2m-2$. The
flow from type III quiver with $N=2m$ to type III quiver with
$N=2m-2$ is given by setting $\vec{x}\rightarrow \vec{x}_1$ and
$\vec{x}_i=\vec{x}_{2m}=\vec{x_1}$ for $i=2,\ldots, 2m-2$.
\\
\indent Notice that the $V$ in \eqref{V3} cannot be obtained from
either \eqref{V1} or \eqref{V2} since both of them have none or only
one single singularities while $V$ in \eqref{V3} has two.
Furthermore, the flow from type II quiver to type I quiver is not
allowed because there is no single singularity in \eqref{V2}, but
there is one in \eqref{V1}. All the flows given above exactly agree
with those obtained from the field theory side. So, we see that the
effect of the $\Omega$ projection is to remove some of the blowing
up, closed string, moduli and therefore the geometry cannot be
completely smoothed out. Generically there remain $\mathbb{ Z}_2$
singularities. Of course higher singularities can be obtained by
bringing together the centers surviving the $\Omega$ projection. We
summarize all possible flows in table \ref{table2}. The UV geometry
is always $AdS_5\times S^5/(\mathbb{Z}_N\times \mathbb{Z}_2)$ with
$\vec{x}_{\textrm{UV}}\rightarrow \infty$. The $V_{\textrm{UV}}$ is
given by that of \eqref{ALEmetric} while $V_{\textrm{IR}}$'s can be
obtained by the $\vec{x}_\textrm{IR}$ given in the table via
\eqref{V1}, \eqref{V2} and \eqref{V3}. In the Flow column, the
notation I($2m+1$)$\rightarrow$ II($2m$) means the flow from type I
quiver with $N=2m+1$ to type II quiver with $N=2m$ etc. The $N=2$,
$USp(2n)_{\textrm{diag}}$ gauge theory is denoted by I(1). The ALE
centers are labeled in the same ordering as in equations \eqref{V1},
\eqref{V2} and \eqref{V3}. Finally, $\vec{x}_{\textrm{IR}}$'s are
the IR points with the notation $\vec{x}'$ denoting any regular
point away from the ALE center $\vec{x}_i$'s.
\TABLE{\begin{tabular}{|c|c|}
  \hline
  Flow                                        & $\vec{x}_{\textrm{IR}}$            \\ \hline
  I($2m+1$) $\rightarrow$ I($1$)              & $\vec{x}_1$                        \\
  II($2m$) $\rightarrow$ I($1$)               & $\vec{x}'$                         \\
  III($2m$) $\rightarrow$ I($1$)              & $\vec{x}_1$                       \\
  I($2m+1$) $\rightarrow$ I($2n+1, n< m$)     & $\vec{x}_i=\vec{x}_1$, $i=2,\ldots , n$  \\
  I($2m+1$) $\rightarrow$ II($2n, n\leq m$)   & $\vec{x}_i=\vec{x}_2$, $i=3,\ldots, n+1$  \\
  II($2m$) $\rightarrow$ II($2n, n<m$)        & $\vec{x}_i=\vec{x}_1$, $i=2,\ldots, n$ \\
  III($2m$) $\rightarrow$ III($2n, n < m$)    & $\vec{x}_i=\vec{x}_1=\vec{x}_{2m}$, $i=2,\ldots, n$ \\
  III($2m$) $\rightarrow$ II($2n, n\leq m-1$) & $\vec{x}_i=\vec{x}_2$, $i=2,\ldots,n$ \\
  III($2m$) $\rightarrow$ I($2n+1, n\leq m-1$)& $\vec{x}_i=\vec{x}_1$, $i=2,\ldots,n$ \\
  \hline
\end{tabular}\caption{All possible RG flows of the $N=2$ quiver gauge
theories arising in type I$'$ theory.}}\label{table2}
\section{Conclusions}\label{conclusion}
We have studied RG flow solutions in the four and two dimensional
field theories on the background of the $A_N$ ALE space. The flows
in two dimensions are similar to the solution given in \cite{rg_int}
with the flat four dimensional space replaced by the ALE space. The
flows are vev flows driven by a vacuum expectation value of a
marginal operator as in the solutions of \cite{rg_int}. The dual
field theory description is that of the (2,0) UV CFT flows to the
(4,0) theory in the IR. The corresponding geometries are
$AdS_3\times S^3/\mathbb{Z}_N$ and $AdS_3\times S^3$. We have
computed the central charges in both the UV and IR to curvature
squared terms in the bulk. The ratio of the central charges to the
leading order contains a factor of $N$ as expected from the ratio of
the volumes of the $S^3$ and $S^3/\mathbb{Z}_N$ on which the six
dimensional supergravity is reduced.  \\
\indent In type IIB theory, we have studied a flow solution
describing an RG flow in four dimensional field theory. It involves
the Green's function on $\textrm{ALE}\times\mathbb{R}^2$, which we
were unable to find explicitely, but whose existence is guaranteed.
The solution interpolates between $AdS_5\times S^5/\mathbb{Z}_N$ and
$AdS_5\times S^5$. The flow is again a vev flow driven by a vacuum
expectation value of a relevant operator of dimension two. The flow
drives the $N=2$ quiver gauge theory with the gauge group $SU(n)^N$
in the UV to the $N=4$ $SU(n)_{\textrm{diag}}$ supersymmetric
Yang-Mills theory in the IR. The hypermultiplets acquire vacuum
expectation values proportional to the identity matrix and break
$SU(n)^N$ to its diagonal subgroup $SU(n)_{\textrm{diag}}$ in the
IR. The central charges $a$ and $c$ have also been computed to the
curvature squared terms.
\\
\indent Moreover, we have studied a flow solution in type I$'$
theory. The flow solution interpolates between $AdS_5\times
S^5/(\mathbb{Z}_N\times \mathbb{Z}_2)$ and $AdS_5\times
S^5/\mathbb{Z}_2$ where the $\mathbb{Z}_2$ is $(-1)^{F_L}\Omega
I_2$. The flow is again driven by a vacuum expectation value of a
relevant operator of dimension two. In contrast to the type IIB
case, the field theory description is more complicated and more
interesting. There are three cases to be considered. For $N$ odd and
$\chi(g,\Omega)=1$, the flow drives the $N=2$ quiver gauge theory
with the gauge group $USp(2n)\times U(2n)\times \ldots\times U(2n)$
to the $N=2$, $USp(2n)_{\textrm{diag}}$ gauge theory. For $N$ even
and $\chi(g,\Omega)=1$, the flow describes an RG flow from $N=2$
quiver $USp(2n)\times U(2n)\times \ldots\times U(2n)\times USp(2n)$
gauge theory to $N=2$, $USp(2n)_{\textrm{diag}}$ gauge theory.
Finally, for $N$ even and $\chi(g,\Omega)=e^{\frac{2\pi i}{N}}$, we
find the flow from $N=2$ quiver $U(2n)\times \ldots \times U(2n)$
gauge theory to $N=2$, $U(2n)_{\textrm{diag}}$ gauge theory for
vanishing expectation values of the antisymmetric bifundamental
scalars. With non-zero antisymmetric scalar expectation values, the
gauge group in the IR is reduced to $USp(2n)_{\textrm{diag}}$.
\\
\indent We have also generalized the previous discussion to RG flows
between two $N=2$ quiver gauge theories in both type IIB and type
I$'$ theories. The gravity solution interpolates between
$AdS_5\times S^5/(\mathbb{Z}_N\times \mathbb{Z}_2)$ and $AdS_5\times
S^5/(\mathbb{Z}_M\times \mathbb{Z}_2)$ geometries. In type IIB
theory, the flows work properly as expected from the field theory
side in a strightforward way. In type I$'$ theory, field theory
considerations forbid some symmetry breaking patterns. However, this
is in agreement with the geometrical picture, after one takes into
account the restrictions put on the geometry by the orientifolding
procedure.
\\
\indent We conclude this paper with a few comments regarding the
type I$'$ case. If we include higher order terms in the effective
action, we need, among other things, to switch on the $F\wedge F$ to
ensure the Bianchi identity for the 5-form
\begin{equation}
d \tilde{F}^{(5)}=\frac{\alpha'}{4}(\textrm{Tr}R\wedge R-
\textrm{Tr}F\wedge F)\delta^{(2)}(\vec{z})
\end{equation}
$F$ being the field strength of the $SO(8)$ gauge group and $\vec z$
a coordinate on the transverse $\mathbb{R}^2$. In particular, we
need to include $SO(8)$ instantons on the ALE spaces (with the
standard metric, the warp factor being irrelevant due to conformal
invariance). It would be interesting to relate ALE's instanton
configurations to the pattern of symmetry breaking of the global
$SO(8)$ group involved in the various flows discussed in the
previous Section. As already mentioned, the UV group is determined
by the holonomy of the flat connection at the ALE's boundary, which
is in turn part of the ADHM data. It would be interesting to
understand in a similar way the IR group.

\acknowledgments This work has been supported in part by the EU
grant UNILHC-Grant Agreement PITN-GA-2009-237920.


\begin{thebibliography}{99}
\bibitem{maldacena} J. M. Maldacena, ``The large $N$ limit of
superconformal field theories and supergravity'', Adv. Theor. Math.
Phys. \textbf{2} (1998) 231-252, arXiv: hep-th/9711200.
\bibitem{fgpw} D.Z.  Freedman, S. Gubser, N. Warner and K. Pilch, ``Renormalization Group Flows from Holography-Supersymmetry and a
c-Theorem'', Adv. Theor. Math. Phys. \textbf{3} (1999), arXiv:
hep-th/9904017.
\bibitem{an} Alexei Khavaev and Nicholas P. Warner, ``A Class of $N=1$ Supersymmetric RG Flows from Five-dimensional $N=8$
Supergravity'', Phys. Lett. \textbf{B495} (2000) 215-222. arXiv:
hep-th/0009159.
\bibitem{gir} L.Girardello, M.Petrini, M.Porrati and A. Zaffaroni, ``Novel Local CFT and exact results on perturbations
of $N=4$ super Yang-Mills from AdS dynamics'', JHEP 12 (1998)
\textbf{022}.
\bibitem{bs} M. Berg and H. Samtleben, ``An exact holographic RG Flow between 2d Conformal Field
Theories'', JHEP 05 (2002) \textbf{006}, arXiv: hep-th/0112154.
\bibitem{gkn} Edi Gava, Parinya Karndumri and K. S. Narain, ``AdS$_3$ Vacua and RG Flows in Three Dimensional Gauged
Supergravities'', JHEP 04 (2010) \textbf{117}, arXiv: 1002.3760.
\bibitem{AP} Auttakit Chatrabhuti and Parinya Karndumri, ``Vacua and
RG flows in $N=9$ three dimensional gauged supergravity'', JHEP 10
(2010) \textbf{098}, arXiv: 1007.5438.
\bibitem{chethan} Chethan Krishnan and Stanislav Kuperstein, ``Gauge
theory RG flows from a warped resolved orbifold'', JHEP 04 (2008)
\textbf{009}, arXiv: 0801.1053.
\bibitem{km} I.R. Klebanov and A. Murugan, "Gauge/Gravity Duality and
Warped Resolved Conifold",  JHEP 03 (2007) \textbf{042}, arXiv:
hep-th/070164.
\bibitem{rg_int} Edi Gava, Parinya Karndumri and K. S. Narain, ``Two
dimensional RG flows and Yang-Mills instantons'', JHEP 03 (2011)
\textbf{106}, arXiv: 1012.4953.
\bibitem{Hawking ALE} G. W. Gibbons and S. W. Hawking,
``Gravitational Multi-instantons'', Phys. Lett. \textbf{B78} (1978)
4.
\bibitem{Massimo} Massimo Bianchi, Francesco Fucito, Giancarlo Rossi
and Maurizio Martellini, ``Explicit Construction of Yang-Mills
Instantons on ALE Spaces'', Nucl. Phys. \textbf{B473} (1996)
367-404, arXiv: hep-th/9601162.
\bibitem{SU2_instOnALE} H. Boutaleb-Joutei, A. Chakrabarti and A.
Comtet, ''Gauge field configurations in curved space-times. IV.
Self-dual $SU(2)$ fields in multicenter spaces'', Phys. Rev.
\textbf{D21} (1980) 2280-2284.
\bibitem{etesi} Gabor Etesi and Tamas Hausel, ``On Yang-Mills
instantons over multi-centered gravitational instantons'', Commun.
Math. Phys. \textbf{235} (2003) 275-288, arXiv: hep-th/0207196.
\bibitem{Sen_K3F_theory} Ashoke Sen, ``F-theory and Orientifolds'', Nucl. Phys. \textbf{B475} (1996)
,562-578, arXiv: hep-th/9605150.
\bibitem{nessi} Nessi Benishti, ``Emerging non-anomalous baryonic
symmetries in the AdS$_5$/CFT$_4$ correspondence'', JHEP 06 (2011)
\textbf{075}, arXiv: 1102.1979.
\bibitem{douglas_quiver} Michael R. Douglas and Gregory Moore,
``D-branes, Quivers, and ALE Instantons'', arXiv: hep-th/9603167.
\bibitem{pol_K3Orientifold} J. Polchinski, ``Tensors from K3
orientifolds'', Phys. Rev. \textbf{D55} (1997) 6423-6428, arXiv:
hep-th/9606165.
\bibitem{nishino} H. Nishino and E. Sezgin, ``New Coupling of
Six-Dimensional Supergravity'', Nucl. Phys. \textbf{B505} (1997)
497, arXiv: hep-th/9703075.
\bibitem{Page} Don N. Page, ``Green's functions for gravitational
multi-instantons'', Phys. Lett. \textbf{B85} (1979) 4.
\bibitem{bergshoeff} E. Bergshoeff, A. Salam and E. Sezgin,
``Supersymmetric $R^2$ action, conformal invariance and the Lorentz
Chern-Simons term in 6 and 10 dimensions'', Nucl. Phys.
\textbf{B279} (1987) 659-683.
\bibitem{pope_massive_3D} H. Lu, C. N. Pope and E. Sezgin, ``Massive
Three-Dimensional Supergravity From $R+R^2$ Action in Six
dimensions'', JHEP 10 (2010) \textbf{016}, arXiv:1007.0173.
\bibitem{kraus} Per Kraus and Finn Larsen, ``Holographic
Gravitational Anomalies'', JHEP 01 (2006) \textbf{022}, arXiv:
hep-th/0508218.
\bibitem{kraus_higherD} P. Kraus and F. Larsen, ``Microscopic black
hole entropy in theories with higher derivatives'', JHEP 09 (2005)
\textbf{034}, arXiv: hep-th/0506176.
\bibitem{lecture_instanton} Stefan Vandoren and Peter van
Nieuwenhuizen, ``Lectures on instantons'', arXiv: 0802.1862.
\bibitem{IIB_schwarz} J. H. Schwarz, ``Covariant Field Equations Of Chiral $N=2$ $D = 10$ Supergravity'', Nucl.
Phys. \textbf{B226}, (1983) 269.
\bibitem{Howe_West} P. Howe and P. C. West, ``The complete $N=2$, $D=10$ supergravity'', Nucl. Phys.
\textbf{B238} (1984) 181.
\bibitem{existence} Chiung-Jue Anna Sung, ``A Note in the Existence
of Positive Green's Function'', Journal of Functional Analysis,
\textbf{156} (1998) 199-207.
\bibitem{etesi_existence} Gabor Etesi and Szilard Szabo, ``Harmonic function and instanton moduli
spaces on the multi-Taub-NUT space'', Commun. Math. Phys.
\textbf{301} (2011) 175-214, arXiv: 0809.0480.
\bibitem{4D_c_a} S. Cremonini, K. Hanaki, J. T. Liu and P. Szepietowski, ``Black holes in five-dimensional gauged
supergravity with higher derivatives'', JHEP 12 (2009) \textbf{045},
arXiv: 0812.3572.
\bibitem{narain} M. Blau, E. Gava and K. S. Narain, ``On subleading
contributions to the AdS/CFT trace anomaly'', JHEP 09 (1999)
\textbf{018}, arXiv: hep-th/9904179.
\bibitem{Anselmi} D. Anselmi and A. Kehagias, ``Subleading
corrections and central charges in the AdS/CFT correspondence'',
Phys. Lett. \textbf{B455} (1999) 155-163, arXiv: hep-th/9812092.
\bibitem{spalinski} A. Fayyazuddin and M. Spalinski, "Large N Superconformal
Gauge Theories and Supergravity Orientifolds" Nucl. Phys.
\textbf{B535} 219 (1998), arXiv: hep-th/9805096.
\bibitem{aharony} O.Aharony, A. Fayyazuddin and J. Maldacena, "The Large Limit
of N=2, N=1 from Three-Branes in F-theory", JHEP 07 (1998)
\textbf{013}, arXiv: hep-th/9806159.
\bibitem{intriligator} J. D. Blum and K. Intriligator ``Consistency conditions for branes at
orbifold singularities'', Nucl. Phys. \textbf{B506} (1997) 223-235,
arXiv: hep-th/9705030.
\end{thebibliography}
\end{document}